\documentclass[runningheads]{llncs}

\usepackage{eccv}

\usepackage{eccvabbrv}

\usepackage{graphicx}
\usepackage{booktabs}

\usepackage{algorithm}
\usepackage{algcompatible}
\algnewcommand\algorithmicreturn{\textbf{return}}
\algnewcommand\RETURN{\State \algorithmicreturn}%
\newcommand{\comm}[1]{\hfill$\triangleright$ {\small \textit{#1}}}

\usepackage{multirow}
\usepackage{colortbl}
\usepackage{xcolor}
\usepackage{enumitem}

\usepackage[accsupp]{axessibility}  

\usepackage{hyperref}

\usepackage{orcidlink}

\begin{document}

\title{Zero-Shot Adaptation for Approximate Posterior Sampling of Diffusion Models in Inverse Problems}

\titlerunning{Zero-Shot Approximate Posterior Sampling}

\author{Ya\c{s}ar Utku Al\c{c}alar\orcidlink{0009-0000-6413-2588} \and
Mehmet Ak\c{c}akaya\orcidlink{0000-0001-6400-7736}}

\authorrunning{Y. U. Al\c{c}alar and M. Ak\c{c}akaya}

\institute{University of Minnesota, Minneapolis\\ 
\email{\{alcal029,akcakaya\}@umn.edu}}

\maketitle

\begin{abstract}

Diffusion models have emerged as powerful generative techniques for solving inverse problems. Despite their success in a variety of inverse problems in imaging, these models require many steps to converge, leading to slow inference time. Recently, there has been a trend in diffusion models for employing sophisticated noise schedules that involve more frequent iterations of timesteps at lower noise levels, thereby improving image generation and convergence speed. However, application of these ideas for solving inverse problems with diffusion models remain challenging, as these noise schedules do not perform well when using empirical tuning for the forward model log-likelihood term weights. To tackle these challenges, we propose zero-shot approximate posterior sampling (ZAPS) that leverages connections to zero-shot physics-driven deep learning. ZAPS fixes the number of sampling steps, and uses zero-shot training with a physics-guided loss function to learn log-likelihood weights at each irregular timestep. We apply ZAPS to the recently proposed diffusion posterior sampling method as baseline, though ZAPS can also be used with other posterior sampling diffusion models. We further approximate the Hessian of the logarithm of the prior using a diagonalization approach with learnable diagonal entries for computational efficiency. These parameters are optimized over a fixed number of epochs with a given computational budget. Our results for various noisy inverse problems, including Gaussian and motion deblurring, inpainting, and super-resolution show that ZAPS reduces inference time, provides robustness to irregular noise schedules and improves reconstruction quality. Code is available at \url{https://github.com/ualcalar17/ZAPS}.

\end{abstract}

\section{Introduction}
\label{sec:intro}

\begin{figure}[tb]
  \centering
  \includegraphics[width=1.0\columnwidth]{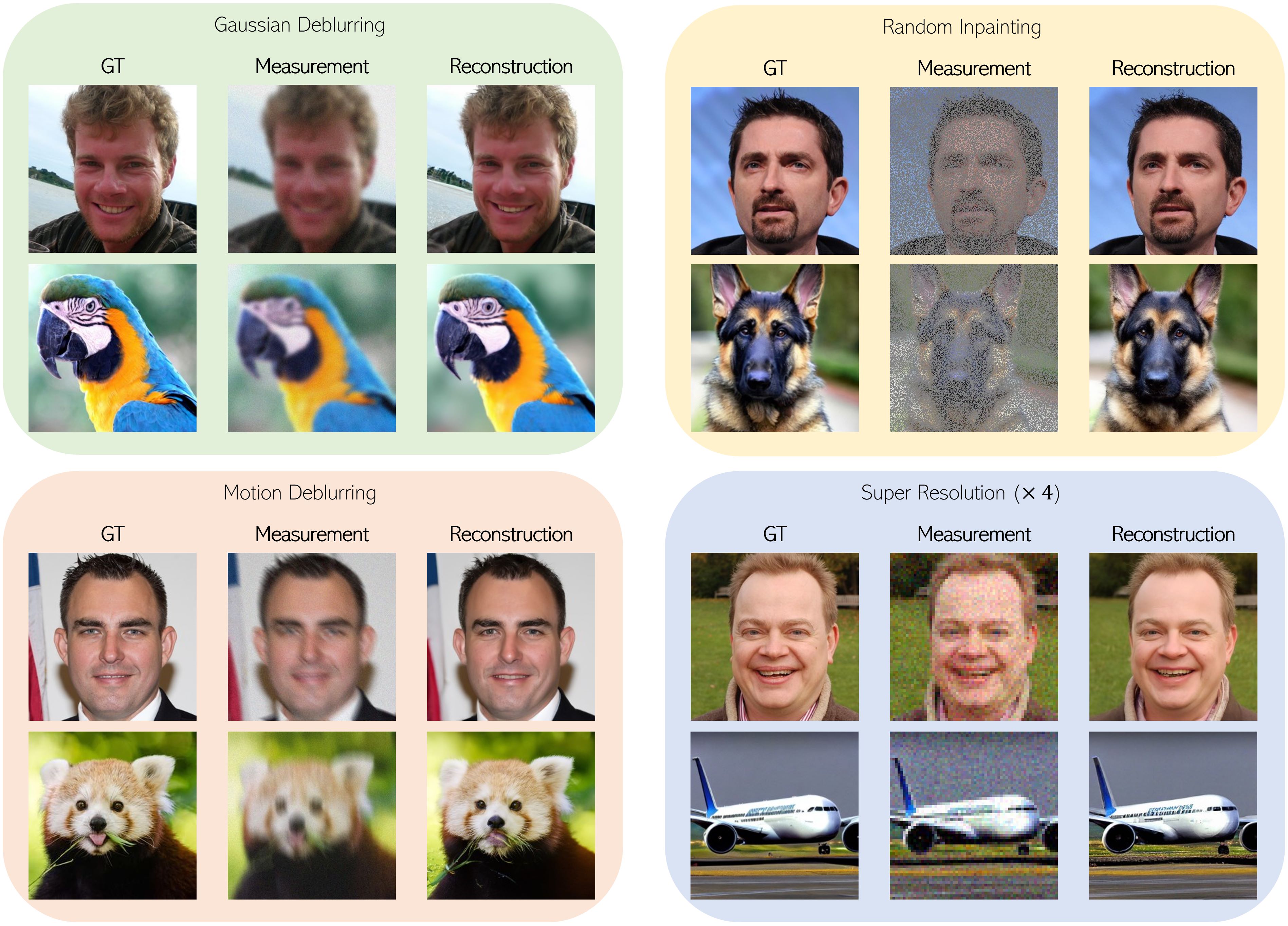}
  \caption{Representative results of our algorithm for four distinct noisy inverse problems ($\sigma=0.05$), showing the ground truth (GT), measurement and reconstruction.}
  \label{fig:1st_fig}
\end{figure}

The forefront of deep generative models is now dominated by diffusion models~\cite{sohl2015deep, song2019scorebased, ho2020ddpm, song2020sde, song2020ddim} in the intricate task of image generation~\cite{dhariwal2021diffusionsweep}. Their capabilities extend across various domains, including computer vision~\cite{baranchuk2021segmentation}, natural language processing~\cite{hoogeboom2021argmax} and temporal data modeling~\cite{alcaraz2022diffusion}. Recently, diffusion models also showed great success in solving noiseless~\cite{chung2022mcg, choi2021ilvr, song2021solvingmedical, song2020sde} and noisy inverse problems~\cite{kawar2022ddrm, chung2022dps, song2022pgdm, song2023latentdiffusion}, owing to their capability to model complicated high-dimensional distributions. Linear inverse problems utilize a known forward model given by
\begin{equation*}
    \mathbf{y}=\mathbf{A}\mathbf{x}_0+\mathbf{n},
\end{equation*}
and aim to deduce the underlying signal/image $\mathbf{x}_0 \in \mathbb{R}^n$ from measurements $\mathbf{y} \in \mathbb{R}^m$, where $\mathbf{n} \in \mathbb{R}^m$ is measurement noise. In practical situations, the forward operator $\mathbf{A}: \mathbb{R}^n \rightarrow \mathbb{R}^m$ is either incomplete or ill-conditioned, necessitating the use of prior information about the signal. Posterior sampling approaches use diffusion models as generative priors and incorporates information from both the data distribution and the forward physics model, allowing for sampling from the posterior distribution $p(\mathbf{x}|\mathbf{y})$ using the given measurement $\mathbf{y}$~\cite{kawar2022ddrm}. In this context, using Bayes' rule, $p(\mathbf{x}|\mathbf{y}) = \frac{p(\mathbf{x})p(\mathbf{y}|\mathbf{x})}{p(\mathbf{y})},$ the problem-specific score is
\begin{equation} 
    \nabla_{\mathbf{x}_t} \log p(\mathbf{x}|\mathbf{y}) = \nabla_{\mathbf{x}_t} \log p(\mathbf{x}) + \nabla_{\mathbf{x}_t} \log p(\mathbf{y}|\mathbf{x}), \label{eq:bayes}
\end{equation}
where $\nabla_{\mathbf{x}_t} \log p(\mathbf{x})$ is approximated via the learned score model $s_\theta(\mathbf{x}_t,t)$. Many of these strategies utilize a plug-and-play (PnP) approach, using a pre-trained unconditional diffusion model as a prior~\cite{cohen2021PnP_inverse,kadkhodaie2021stochastic,chan2016PnP_ADMM, graikos2022PnPriors, laumont2022bayesianPnP, tumanyan2023plug}, and integrate the forward model during inference to address various inverse problem tasks.

The complexity for these approaches arises in obtaining the latter forward model log-likelihood term in \cref{eq:bayes}, which guides the diffusion to a target class~\cite{sohl2015deep, dhariwal2021diffusionsweep}. While exact calculation is intractable, several approaches have been proposed to approximate this term. Among these, RED-diff~\cite{mardani2023red_diff} employs a variational sampler that uses a combination of measurement consistency loss and score matching regularization. Another technique, DSG~\cite{yang2024guidance}, uses a spherical Gaussian constraint for denoising steps, allowing for larger step sizes. A class of methods utilize projections onto the convex measurement subspace after the unconditional update through score model~\cite{chung2022ccdf, choi2021ilvr, song2020sde}. Although these projections improve consistency between measurements and the sample, they are noted to lead to artifacts, such as boundary effects~\cite{chung2022mcg}. Thus, more recent approaches aimed to approximate the log-likelihood term in \cref{eq:bayes} different ways. Noting
\begin{equation}
    p_t(\mathbf{y}|\mathbf{x}_t) = \int_{\mathbf{x}_0} p(\mathbf{x}_0|\mathbf{x}_t) p(\mathbf{y}|\mathbf{x}_0)\mathbf{dx}_0, \label{eq:int}
\end{equation}
DPS~\cite{chung2022dps} uses the posterior mean $\mathbf{\hat{x}}_0 = \mathbf{\hat{x}}_0({\bf x}_t) \triangleq \mathbb{E}[{\bf x}_0 | {\bf x}_t] = \mathbb{E}_{\mathbf{x}_0 \sim p(\mathbf{x}_0|\mathbf{x}_t)} [\mathbf{x}_0]$, 
to approximate $p(\mathbf{y}|\mathbf{x}_t) = \mathbb{E}_{\mathbf{x}_0 \sim p(\mathbf{x}_0|\mathbf{x}_t)}[p(\mathbf{y}|\mathbf{x}_0)]$ as
\begin{equation*}
	p(\mathbf{y}|\mathbf{x}_t) = \mathbb{E}_{\mathbf{x}_0 \sim p(\mathbf{x}_0|\mathbf{x}_t)}[p(\mathbf{y}|\mathbf{x}_0)] \simeq  p\Big(\mathbf{y} | \mathbb{E}_{\mathbf{x}_0 \sim p(\mathbf{x}_0|\mathbf{x}_t)} [\mathbf{x}_0]\Big) = p(\mathbf{y} | \hat{\bf x}_0) .
\end{equation*}
Another technique, $\mathrm{\Pi}$GDM~\cite{song2022pgdm} approximates \cref{eq:int} as a Gaussian centered around ${\bf A\hat{x}_0}$
\begin{equation}
    \int_{\mathbf{x}_0} p(\mathbf{x}_0|\mathbf{x}_t) p(\mathbf{y}|\mathbf{x}_0)\mathbf{dx}_0 \simeq \mathcal{N}(\mathbf{A}\hat{\mathbf{x}}_0, r_t^2\mathbf{AA}^\top + \sigma_y^2 \mathbf{I}),  \label{eq:pgdm_integral}
\end{equation}
and uses it for guidance. In these works, log-likelihood weights (or gradient step sizes), $\{\zeta_t\}$ are introduced to further control the reconstruction as
\begin{equation} 
	\nabla_{\mathbf{x}_t} \log p(\mathbf{x}|\mathbf{y}) = \nabla_{\mathbf{x}_t} \log p(\mathbf{x}) + \zeta_t \nabla_{\mathbf{x}_t} \log p(\mathbf{y}|\mathbf{x}). \label{eq:gradsize}
\end{equation}
\looseness=-1
While DPS demonstrates high performance in various inverse problem tasks, it suffers from the drawback of requiring a large number of sampling steps, resulting in prolonged reconstruction time. $\mathrm{\Pi}$GDM accelerates this process by adopting regular (linear) jumps approach across the schedule. However, utilizing more complicated schedules, where the jumps are irregular introduces a challenge, as it requires distinct log-likelihood weights, $\zeta_t$, for each timestep. Heuristic adjustment of these weights is difficult and frequently leads to undesirable outcomes.
In this work, by taking an inspiration from zero-shot/test-time self-supervised models~\cite{sun2020test,yaman2021zero} we propose to learn the log-likelihood weights for a fixed number of sampling steps and fine-tune them over a few epochs. It is crucial to note that fine-tuning DPS (or $\mathrm{\Pi}$GDM) entails saving computational graphs for each unroll, leading to memory issues and slow backpropagation. Thus, we also propose to approximate the Hessian of the data probability using a wavelet-based diagonalization strategy~\cite{ghael1997diagonalization}, and learn these diagonal values for each timestep as well. \cref{fig:1st_fig} shows representative results for our method. Our key contributions include:
\begin{itemize}[label=$\bullet$]
  \item We introduce zero-shot approximate posterior sampling (ZAPS), leveraging zero-shot learning for dynamic automated hyperparameter tuning in the inference phase to improve solution of noisy inverse problems via diffusion models. This method fortifies the robustness of the sampling process, attaining a state-of-the-art performance~\cite{chung2022dps,song2022pgdm,kawar2022ddrm} in sampling outcomes. To the best of our knowledge, our method is the first attempt to learn the log-likelihood weights for solving inverse problems via diffusion models by using a measurement-consistent loss when the sampling noise schedule consists of irregular jumps across timesteps.
  \item We provide a well-designed approximation for the Hessian of the logarithm of the prior, enabling a computationally efficient and trainable posterior computation.
  \item We showcase the efficacy of incorporating a learnable log-likelihood weights for each diffusion step during the reverse diffusion process through both quantitative and qualitative assessments on FFHQ and ImageNet datasets. Our approach not only outperforms state-of-the-art, but it also substantially reduces the required number of sampling steps from 1000 to $\sim$20-to-30, facilitating convergence with fewer total neural function evaluations (NFEs).
\end{itemize}

\section{Related Works}
\label{sec:related_work}
\subsubsection{Diffusion Models.}
During training, diffusion models~\cite{ho2020ddpm,song2020sde} add Gaussian noise to an image with a fixed increasing variance schedule, e.g. linear or exponential, $\beta_1,\beta_2,...,\beta_T$ until pure noise is obtained, and learns a reverse diffusion process, where a neural network is trained to gradually remove noise and reconstruct the original image. Let $\mathbf{x}_0 \sim p_{\text{data}}(x)$ be samples from the data distribution, and $\mathbf{x}_{\{1:T\}} \in \mathbb{R}^d$ be noisy latent variables. By taking $\alpha_t=1-\beta_t$ and $\bar{\alpha}_t=\prod_{s=1}^{t} \alpha_s$, the Markovian forward process can be written as
\begin{equation}
    q(\mathbf{x}_t | \mathbf{x}_0) = \mathcal{N}(\mathbf{x}_t | \sqrt{\bar{\alpha}_t} \mathbf{x}_0, (1-\bar{\alpha}_t)\mathbf{I}). \label{eq:ddpm_forward}
\end{equation}
By using the reparameterization trick and \cref{eq:ddpm_forward}, $\mathbf{x}_t$ can be sampled as
\begin{equation}
    \mathbf{x}_t(\mathbf{x}_0,\boldsymbol{\epsilon}) = \sqrt{\bar{\alpha}_t}\mathbf{x}_0+ \sqrt{1-\bar{\alpha}_t}\boldsymbol{\epsilon} \quad \text{where} \quad \boldsymbol{\epsilon} \sim \mathcal{N}(\boldsymbol{\epsilon};0, \mathbf{I}). \label{eq:repara_ddpm}
\end{equation}
Consequently, denoising diffusion probabilistic models (DDPMs)~\cite{ho2020ddpm} learns the reverse process by minimizing a lower bound on the log prior via:
\begin{equation}
    L_t(\theta) = \mathbb{E}_{t, \mathbf{x}_0, \boldsymbol{\epsilon}} \left\| \boldsymbol{\epsilon} - \boldsymbol{\epsilon}_\theta(\mathbf{x}_t(\mathbf{x}_0, \boldsymbol{\epsilon}), t) \right\|_2^2. \label{eq:eps_score}
\end{equation}
Furhtermore, it can be shown that epsilon matching in \cref{eq:eps_score} is analogous to the denoising score matching (DSM)~\cite{vincent2011scorematching,song2019scorebased} objective up to a constant:
\begin{equation}
    \min_\theta \mathbb{E}_{\mathbf{x}_t, \mathbf{x}_0, \epsilon} \left\| \mathbf{s}_\theta(\mathbf{x}_t,t) - \nabla_{\mathbf{x}_t} \log q(\mathbf{x}_t | \mathbf{x}_0) \right\|_2^2,
\end{equation}
in which $\mathbf{s}_\theta(\mathbf{x}_t,t) = -\frac{\boldsymbol{\epsilon}_\theta(\mathbf{x}_t,t)}{\sqrt{1-\bar{\alpha}_t}}$. Using Tweedie's formula and \cref{eq:repara_ddpm}, posterior mean for $p(\mathbf{x}_0|\mathbf{x}_t)$ can be found as:
\begin{equation}
    \mathbf{\hat{x}}_0 = \frac{1}{\sqrt{\bar{\alpha}_t}}\Big(\mathbf{x}_t + (1-\bar{\alpha}_t)\mathbf{s}_\theta(\mathbf{x}_t,t)\Big). \label{eq:tweedie_denoise}
\end{equation}
Sampling $\mathbf{x}_{t+1}$ from $p(\mathbf{x}_{t+1}|\mathbf{x}_t)$ can be done using ancestral sampling by iteratively computing:
\begin{equation}
    \mathbf{x}_{t-1} = \frac{1}{\sqrt{\alpha_t}} \left( \mathbf{x}_{t-1} - \frac{1-\alpha_t} {\sqrt{1-\bar{\alpha}_t}} \boldsymbol{\epsilon}_\theta(\mathbf{x}_t,t) \right) + \sigma_t \mathbf{z},
\end{equation}
where $\mathbf{z} \sim \mathcal{N}(0, \mathbf{I})$ and $\sigma_t^2=\tilde{\beta}_t=\frac{1 - \bar{\alpha}_{t-1}}{1 - \bar{\alpha}_t} \beta_t$. It is also worth noting that the DDPM is equivalent to the variance preserving stochastic differential equations (VP-SDEs)~\cite{song2020sde}. 

\subsubsection{Solving Inverse Problems via Diffusion Models.}
When solving inverse problems via diffusion models, the main challenge is to find an approximation to the log-likelihood term, $\nabla_{\mathbf{x}_t} \log p(\mathbf{y}|\mathbf{x}),$ as discussed earlier. One recent method, denoising diffusion restoration models (DDRM)~\cite{kawar2022ddrm}, utilizes a spectral domain approach, allowing the incorporation of noise from the measurement domain into the spectral domain through singular value decomposition (SVD). However, the application of SVD is computationally expensive~\cite{chung2022dps}. Manifold Constrained Gradient (MCG)~\cite{chung2022mcg} method applies projections after the MCG correction as:
\vspace{-11pt}

\begin{align}
    \mathbf{x}_{t-1}^{'} &= f(\mathbf{x}_t, \mathbf{s}_\theta) - \zeta \nabla_{\mathbf{x}_t} \|\mathbf{K}(\mathbf{y}-\mathbf{A}\hat{\mathbf{x}}_0)\|_2^2 + g(\mathbf{x}_t)\mathbf{z}, \quad \mathbf{z} \sim \mathcal{N}(0,\mathbf{I}), \label{eq:mcg_update}\\
    \mathbf{x}_{t-1} &= \mathbf{Hx}_{t-1} + \mathbf{b},
\end{align}
where $\zeta$ and $\mathbf{H}$ are dependent on noise covariance. MCG update of \cref{eq:mcg_update} projects estimates onto the measurement subspace, thus they may fall off from the data manifold~\cite{chung2022dps}. Hence, DPS proposes to update without projections as:
\begin{equation}
    \mathbf{x}_{t-1} = \mathbf{x}^{'}_{t-1} - \zeta_t \nabla_{\mathbf{x}_t} \|\mathbf{y}-\mathbf{A}\hat{\mathbf{x}}_0\|_2^2, \label{eq:dps_update}
\end{equation}
Note \cref{eq:dps_update} is equivalent to \cref{eq:mcg_update} when $\mathbf{K=I}$, and it reduces to the following when the forward operator is linear:
\begin{equation}
    \mathbf{x}_{t-1} = \mathbf{x}^{'}_{t-1} + \zeta_t \frac{\partial \hat{\mathbf{x}}_0}{\partial \mathbf{x}_t} \mathbf{A}^\top (\mathbf{y}-\mathbf{A}\hat{\mathbf{x}}_0) \label{eq:liner_dps_update}
\end{equation}
$\mathrm{\Pi}$GDM~\cite{song2022pgdm}, on the other hand, utilizes a Gaussian centered around $\mathbf{\hat{x}}_0$ that is defined in \cref{eq:tweedie_denoise} to obtain the following score approximation:
\begin{equation}
	\nabla_{\mathbf{x}_t} \log p_t(\mathbf{y}|\mathbf{x}_t) \simeq  \frac{\partial \hat{\mathbf{x}}_0}{\partial \mathbf{x}_t} \mathbf{A}^\top (r_t^2 \mathbf{A} \mathbf{A}^\top + \sigma_y^2 \mathbf{I})^{-1} (\mathbf{y} - \mathbf{A}\hat{\mathbf{x}}_0) . \label{eq:pgdm_update}
\end{equation}
In cases where there is no measurement noise $(\sigma_y=0)$, \cref{eq:pgdm_update} simplifies to:
\begin{equation}
    \nabla_{\mathbf{x}_t} \log p_t(\mathbf{y}|\mathbf{x}_t) \simeq r_t^{-2} \frac{\partial \hat{\mathbf{x}}_0}{\partial \mathbf{x}_t} \mathbf{A}^\dagger (\mathbf{y}-\mathbf{A}\hat{\mathbf{x}}_0)
\end{equation}
where $\mathbf{A}^\dagger$ denotes the Moore-Penrose pseudoinverse of $\mathbf{A}$. We note that using Woodbury matrix identity (derived in SuppMat), one can simplify \cref{eq:pgdm_update} to:
\begin{equation}
    \nabla_{\mathbf{x}_t} \log p_t(\mathbf{y}|\mathbf{x}_t) \simeq \frac{\partial \hat{\mathbf{x}}_0}{\partial \mathbf{x}_t} (\mathbf{A}^\top \mathbf{A} + \eta \mathbf{I})^{-1} \mathbf{A}^\top (\mathbf{y}-\mathbf{A}\hat{\mathbf{x}}_0), \quad \text{where } \eta=\frac{\sigma_y^2}{r_t^2}. \label{eq:pgdm_update_woodbury}
\end{equation}
From \cref{eq:pgdm_update_woodbury}, the similarity between DPS and $\mathrm{\Pi}$GDM updates can be seen, with $(\mathbf{A}^\top \mathbf{A} + \eta \mathbf{I})^{-1}$ term being the difference. Note the DPS update in~\cref{eq:dps_update} works with non-linear operators, while $\mathrm{\Pi}$GDM's update does not rely on the differentiability of the forward operator, as long as a pseudo-inverse-like operation can be derived.

\subsubsection{Improved Irregular Noise Schedules for Image Generation.}
Diffusion models typically utilize well-defined fixed noise schedules, with examples including linear or exponential ones. Lately, more sophisticated methods have been developed that sweep across these schedules and take samples in irregular timesteps~\cite{dhariwal2021diffusionsweep,karras2022elucidating} for unconditional image generation. The idea behind this strategy hinges on more frequent sampling for lower noise levels, making it possible to use considerably less number of sampling steps.

Most of the aforementioned studies that solve inverse problems via diffusion models used the same number of steps that the unconditional diffusion model was trained for~\cite{song2020sde,chung2022mcg,chung2022dps}. Nonetheless, there has been a notable trend favoring shorter schedules characterized by linear jumps for inverse problems, where the log-likelihood weights were hand-tuned by trial-and-error~\cite{song2022pgdm,mardani2023red_diff} when using reduced number of steps.
While these approaches have proven effective, they still require a large number of sampling steps or heuristic tuning of the log-likelihood weights, $\{\zeta_t\}$ in \cref{eq:gradsize} to achieve good performance. The former issue leads to lengthy and potentially impractical computational times, while the latter issue results in generalizability difficulties for adoption at different measurement noise levels and variations in the measurement operators. Furthermore, the irregular jump strategy that has been powerful for image generation has not garnered significant attention for inverse problems, mainly due to the impracticality of empirically tuning the log-likelihood weights. Thus, a method that automatically selects and adjusts log-likelihood weights based on the provided measurements for arbitrary noise schedules, instead of requiring manual tuning, holds significant potential for improving robustness and image quality.

\section{Methodology}
\label{sec:methodology}

\subsection{Zero-shot Fine Tuning of Log-Likelihood Weights}

\begin{figure}[tb]
  \centering
  \includegraphics[width=1.0\columnwidth]{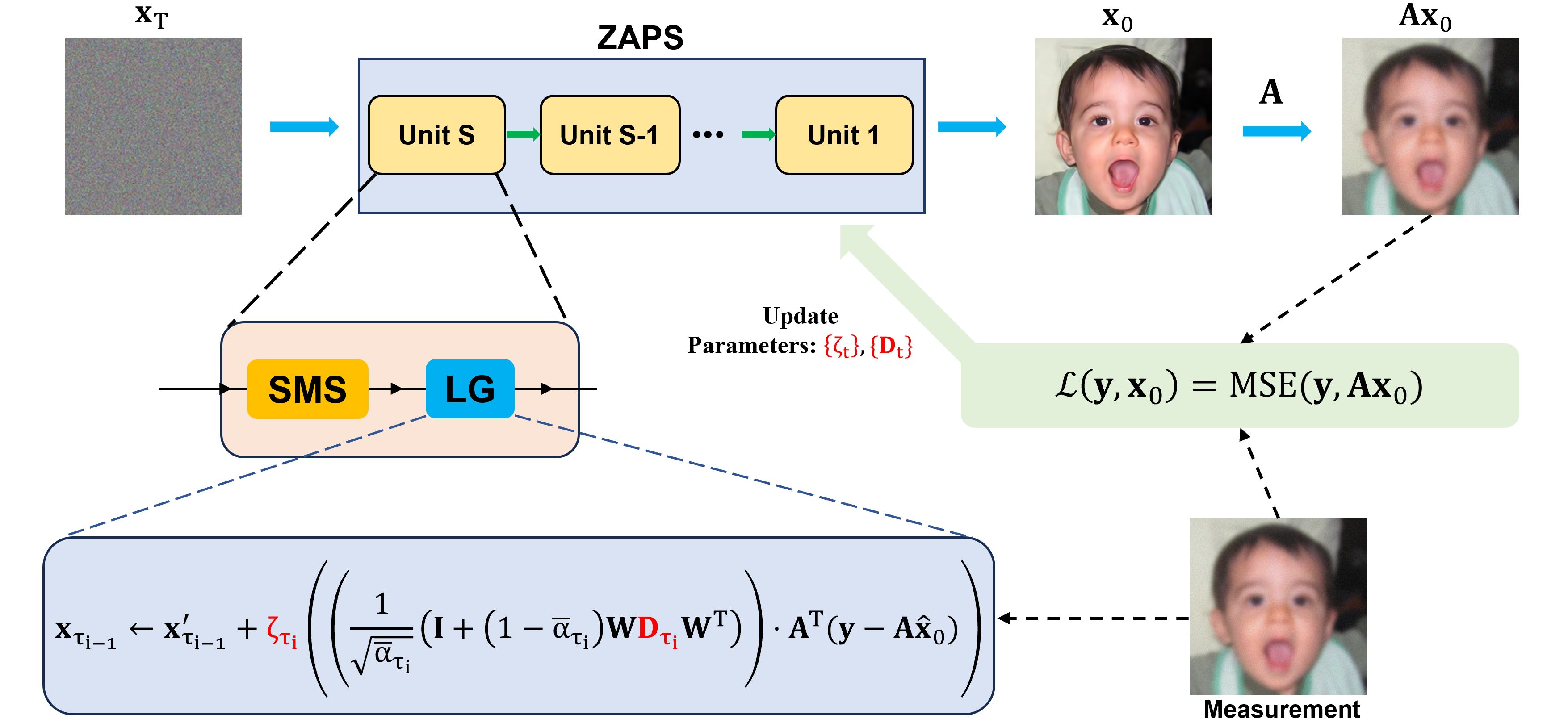}
  \caption{Our zero-shot approximate posterior sampling (ZAPS) approach unrolls the sampling process for a fixed number of $S$ steps for arbitrary/irregular noise schedules, alternating between score model sampling (SMS) and likelihood guidance (LG). Our zero-shot fine-tuning approach has two key components: 1) The Hessian of the log prior is approximated using a discrete wavelet transform diagonalization technique, 2) Both the diagonal matrices, $\{{\bf D}_t\}$ and the log-likelihood weights, $\{\zeta_t\}$ are updated during fine-tuning. The fine-tuning is done for a fixed number of epochs with a given NFE budget, yielding a faster and more robust \emph{adaptive} inverse problem solver.}
  \label{fig:2nd_fig}
\end{figure}

In this work, we propose a robust automated approach for setting the log-likelihood weights at each timestep for arbitrary noise sampling schedules to improve posterior sampling with the given measurements during inference. This allows for a stable reconstruction for different sweeps across noise schedules. Furthermore, the weights themselves are image-specific, which improves the performance compared to the former approaches. For estimating the likelihood in \cref{eq:bayes}, we use the update in DPS~\cite{chung2022dps}:
\begin{align}
	\nabla_{\mathbf{x}_t} \log p(\mathbf{y}|\mathbf{x}_t) \simeq
	\nabla_{\mathbf{x}_t} \|\mathbf{y}-\mathbf{A}\hat{\mathbf{x}}_0\|_2^2 = -
	\frac{\partial \hat{\mathbf{x}}_0}{\partial \mathbf{x}_t} \mathbf{A}^\top (\mathbf{y}-\mathbf{A}\hat{\mathbf{x}}_0),
\end{align}
although as noted before, the $\mathrm{\Pi}$GDM~\cite{song2022pgdm} update in \cref{eq:pgdm_update_woodbury} is also similar. Thus we emphasize that while we chose DPS as baseline for its versatility in inverse problems, our ZAPS strategy is applicable to other diffusion models for inverse problems. Recalling the definition of $\mathbf{\hat{x}}_0$ in \cref{eq:tweedie_denoise}, we note
\begin{equation}
\frac{\partial \hat{\mathbf{x}}_0}{\partial \mathbf{x}_t} = \frac{1}{\sqrt{\bar{\alpha}_t}} \left( \mathbf{I}+(1-\bar{\alpha}_t)\frac{\partial \mathbf{s}_\theta(\mathbf{x}_t,t)}{\partial \mathbf{x}_t}\right). \label{eq:hessian1}
\end{equation} 
Thus, ignoring the calculation and storage of the matrix $\frac{\partial \mathbf{s}_\theta(\mathbf{x}_t,t)}{\partial \mathbf{x}_t}$ for now, one needs to fine tune the log-likelihood weights $\{\zeta_t\}$ in
\begin{equation} \label{eq:loglikelihoodunroll}
	\nabla_{\mathbf{x}_t} \log p(\mathbf{x}) + \zeta_t \frac{1}{\sqrt{\bar{\alpha}_t}} \left( \mathbf{I}+(1-\bar{\alpha}_t)\frac{\partial \mathbf{s}_\theta(\mathbf{x}_t,t)}{\partial \mathbf{x}_t}\right)\mathbf{A}^\top (\mathbf{y}-\mathbf{A}\hat{\mathbf{x}}_0).
\end{equation} 
This is done based on the concept of algorithm unrolling~\cite{gregor2010learning, knoll2020SPM, hammernik2023physics} in physics-driven deep learning by fixing the number of sampling steps $T$. Then the whole posterior sampling process is described as alternating between DDPM sampling using the pre-trained unconditional score model, followed by the log-likelihood term guidance in \cref{eq:loglikelihoodunroll} for $T$ steps. This ``unrolled'' network is fine-tuned end-to-end, where the only updates are made to $\{\zeta_t\}$ and no fine-tuning is performed on the unconditional score function, $\mathbf{s}_\theta(\mathbf{x}_t,t)$. This also alleviates the need for backpropagation across the score function network, leading to further savings in computational time. The fine-tuning is performed using a physics-inspired loss function that evaluates the consistency of the final estimate and the measurements: $\mathcal{L}(\mathbf{y},\mathbf{x}_0) = ||{\bf y} - {\bf A}{\bf x}_0||_2^2.$ High-level explanation for our algorithm is given in \cref{fig:2nd_fig}.

\subsection{Approximation for the Hessian of the Log Prior}

\begin{algorithm}[tb]
  \caption{ZAPS: Zero-Shot Approximate Posterior Sampling}
  \begin{algorithmic}[1]
    \REQUIRE $T, \text{ } \mathbf{y}, \text{ }\{\Tilde{\sigma_i}\}_{i=1}^{T}$, orthogonal DWT ({\bf W})
    \STATE $\mathbf{x}_T \sim \mathcal{N}\mathbf{(0, I)}$ 
    \STATE $\tau \subset [1,...,T]$ extending over a length of $S<T$
    \FOR {epoch $\textbf{in}\text{ range(epochs)}$}
        \FOR {$i = S,...,1$}
            \STATE $\mathbf{\hat{s}} \gets \mathbf{s}_\theta(\mathbf{x}_{\tau_i},\tau_i)$ \comm{Score computation}
            \vspace{1ex}
            \STATE $\mathbf{\hat{x}}_0 \gets \frac{1}{\sqrt{\bar{\alpha}_{\tau_i}}}(\mathbf{x}_{\tau_i} + (1-\bar{\alpha}_{\tau_i})\mathbf{\hat{s}})$ \comm{Tweedie denoising}
            \vspace{1ex}
            \STATE $\mathbf{z} \sim \mathcal{N}\mathbf{(0, I)} \text{ if } \tau_i>1$, else $\mathbf{z=0}$ 
            \vspace{1ex}
            \STATE $\mathbf{x}^{'}_{{\tau_{i-1}}} \gets \frac{\sqrt{\alpha_{\tau_i}}(1-\bar{\alpha}_{\tau_{i-1}})}{1-\bar{\alpha}_{\tau_i}} \mathbf{x}_{\tau_i} + \frac{\sqrt{\bar{\alpha}_{\tau_{i-1}}}\beta_{\tau_i}}{1-\bar{\alpha}_{\tau_i}} \mathbf{\hat{x}}_0 + \tilde{\sigma}_{\tau_i} \mathbf{z}$
            \vspace{1ex}
            \STATE \textcolor{blue}{$\mathbf{x}_{\tau_{i-1}} \gets \mathbf{x}^{'}_{\tau_{i-1}} + \zeta_{\tau_i} \left(\left(\frac{1}{\sqrt{\bar{\alpha}_{\tau_i}}}\Big(\mathbf{I}+(1-\bar{\alpha}_{\tau_i})\mathbf{WD}_{\tau_i}\mathbf{W}^\top \Big)\right) \cdot \mathbf{A}^\top (\mathbf{y}-\mathbf{A\hat{x}}_0)\right)$}
        \ENDFOR
        \STATE \textcolor{blue}{Update network parameters $\{\zeta_t\}$ and $\{\mathbf{D}_t\}$}
    \ENDFOR
    \RETURN{} $\mathbf{x}_0$
  \end{algorithmic}
  \label{algo:alg1}
\end{algorithm}

Implementing the zero-shot update for \cref{eq:loglikelihoodunroll} poses various challenges, since backpropagation through the unrolled network to update all $\{\zeta_t\}$ requires another backpropagation through the Jacobian of the score function at each time step. This can only be done by retaining the computational graphs that are created when calculating the Jacobian term in \cref{eq:loglikelihoodunroll}, which quickly explodes memory requirements, especially when the number of sampling steps increases. Also, backpropagating through multiple graphs at the end to only update the log-likelihood weights is time-inefficient and causes prolonged sampling times. Hence, we propose to approximate the Jacobian using inspirations from wavelet-based signal processing techniques and propose to learn this approximation to improve the overall outcome. Noting that $\mathbf{s}_\theta(\mathbf{x}_t,t)$ in \cref{eq:hessian1} is an approximation of the log-gradient of the true prior $p({\bf x})$, we have
\begin{equation}
    \frac{\partial \hat{\mathbf{x}}_0}{\partial \mathbf{x}_t} = \frac{1}{\sqrt{\bar{\alpha}_t}} \left( \mathbf{I}+(1-\bar{\alpha}_t)\frac{\partial^2 \log p_t(\mathbf{x}_t)}{\partial \mathbf{x}_t^2}\right). \label{eq:hessian}
\end{equation}
In order to make a backpropagation to update these weights, one needs to calculate the Hessian matrix, $\frac{\partial^2 \log p_t(\mathbf{x}_t)}{\partial \mathbf{x}_t^2}$ given in \cref{eq:hessian}. This matrix is the negative of the observed Fisher information matrix, whose expected value is the Fisher information matrix. It is also known that in the limit, it approximates the inverse covariance matrix of the maximum likelihood estimator. Furthermore, under mild assumptions about continuity of the prior, the observed Fisher information matrix is symmetric. Thus, an appropriate decorrelating unitary matrix can be used to diagonalize it. While finding the desired unitary matrix is equally time-consuming as calculating this Hessian, several pre-determined unitary transforms have been proposed for decorrelation in the signal processing community for different applications \cite{ghael1997diagonalization, taam1987approximate, qu2004using}. 
Of particular note is the use of unitary wavelet transforms for Wiener filtering \cite{ghael1997diagonalization}, where these transforms were utilized for their tendency to decorrelate data, i.e. approximate the Karhunen-Loeve transform~\cite{qu2004using}. In this work, we also use these decorrelating properties to approximately diagonalize the Hessian of the log prior, $\frac{\partial^2 \log p_t(\mathbf{x}_t)}{\partial \mathbf{x}_t^2}$ using fixed orthogonal discrete wavelet transforms (DWT): 
\begin{equation}
	\frac{\partial^2 \log p_t(\mathbf{x}_t)}{\partial \mathbf{x}_t^2} \simeq \mathbf{WD}_t\mathbf{W}^{\top}, \label{eq:hessian_approx}
\end{equation}
where $\mathbf{W}$ is an orthogonal DWT. By making this approximation, backpropagation through the score model can also be avoided, and only the diagonal values in distinct $\{\mathbf{D}_t\}$ matrices needs to be learned. Our final algorithm to sample from pure noise with fine-tuning is given in \cref{algo:alg1}.

\section{Evaluation}
\subsection{Experimental Setup and Implementation Details}

We comprehensively evaluated our method, examining its performance through both qualitative and quantitative analyses using FFHQ~\cite{Karras2019ffhq} and ImageNet~\cite{deng2009imagenet} datasets with size $256\times256\times3$. Pre-trained unconditional diffusion models trained on FFHQ and ImageNet were taken from \cite{choi2021ilvr} and \cite{dhariwal2021diffusionsweep} respectively, and used without retraining. For our experiments, we sampled 1000 images from FFHQ and ImageNet validation sets. All images underwent pre-processing to be normalized in the range $[0,1]$. During all the evaluations, a Gaussian measurement noise with $\sigma=0.05$ was used. For the orthogonal DWT, Daubechies 4 wavelet was utilized. For our quantitative evaluations, we employed 30 sampling steps with a schedule of "15,10,5", and 10 epochs for fine-tuning, resulting in a total of 300 NFEs. As noted in \cite{dhariwal2021diffusionsweep}, superior schedules may exist but it requires substantial computational time to try out all possible schedules. Thus, we opted a schedule that is simple, and samples more frequently at the lower noise levels~\cite{dhariwal2021diffusionsweep}. More details about the network architectures and hyperparameter choices are given in SuppMat.

\begin{figure}[tb]
  \centering
  \includegraphics[width=\columnwidth]{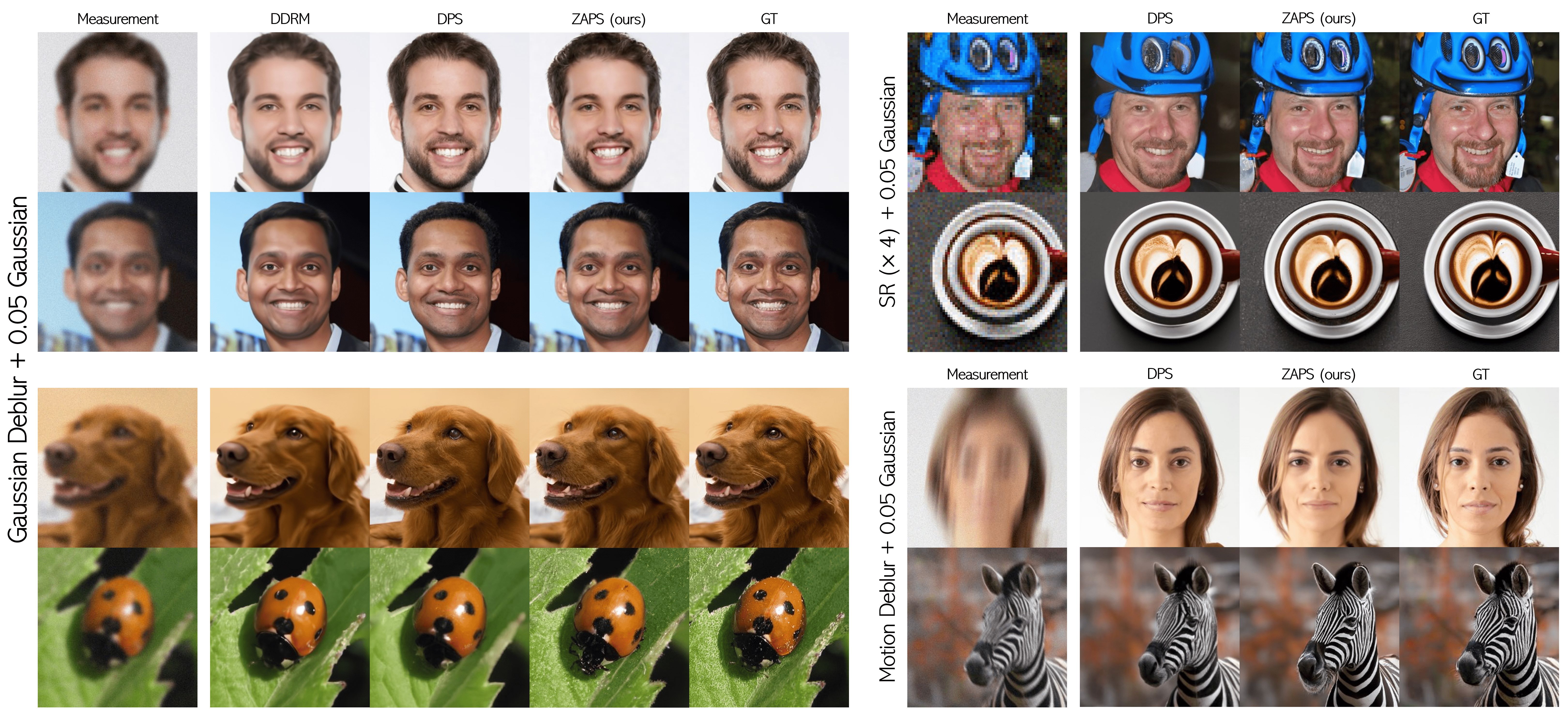}
  \vspace{-.5cm}
  \caption{Representative images using various methods for solving Gaussian deblurring, motion deblurring and super-resolution ($\times4$) tasks. Proposed method qualitatively improves upon each method, including the baseline state-of-the-art DPS.}
  \label{fig:3}
\end{figure}

\begin{figure}[tb]
  \centering
  \includegraphics[width=\columnwidth]{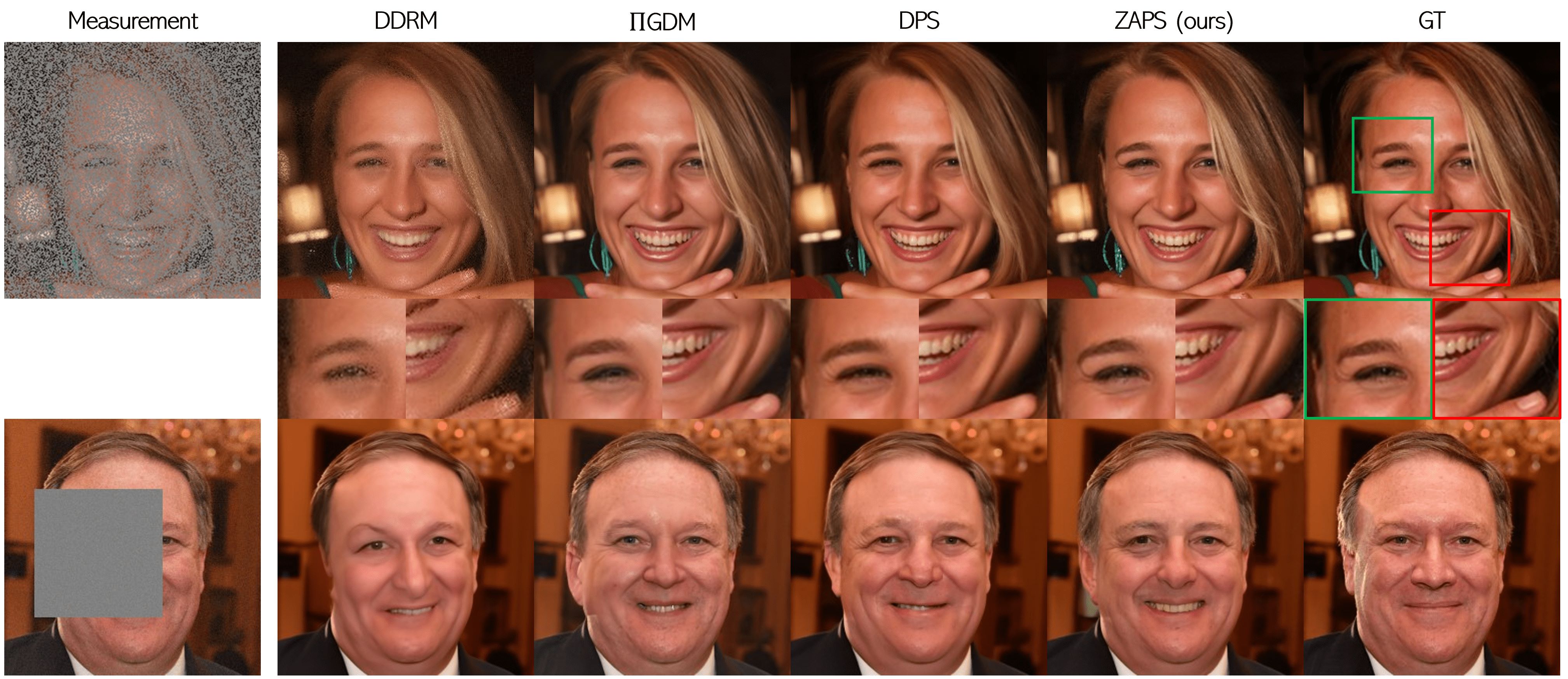}
  \caption{Illustrative images using state-of-the-art methods for random ($70\%$) and box ($128\times128$) inpainting. Proposed method improves upon DDRM, while achieving similar performance to $\mathrm{\Pi}$GDM and DPS, with subtle improvements shown in zoomed insets.}
  \label{fig:4}
\end{figure}

\subsection{Experiments on Linear Inverse Problems}
\subsubsection{Problem Setup.}
We focused on the following linear inverse problems: (1) Gaussian deblurring, (2) inpainting, (3) motion deblurring, (4) super-resolution. For Gaussian deblurring, we considered a kernel of size $61\times61$ with a standard deviation $\sigma=3.0$. For inpainting, we considered two different scenarios wherein we randomly masked out $70\%$ and a $128\times128$ box region of the image, applied uniformly across all three channels. For motion blur, we generated the blur kernel via the code\footnote{\url{https://github.com/LeviBorodenko/motionblur}}, with $61\times61$ kernel size and $0.5$ intensity, as in~\cite{chung2022dps}. Finally, for super-resolution, we considered bicubic downsampling. All measurements are obtained through applying the forward model to the ground truth image.

\vspace{-.2cm}

\subsubsection{Comparison Methods.}
We compared our method with score-SDE~\cite{choi2021ilvr,chung2022ccdf,song2020sde}, manifold constrained gradients (MCG)~\cite{chung2022mcg}, denoising diffusion restoration models (DDRM)~\cite{kawar2022ddrm}, diffusion posterior sampling (DPS)~\cite{chung2022dps} and pseudo-inverse guided diffusion models ($\mathrm{\Pi}$GDM)~\cite{song2022pgdm}. We note that our implementation of score-SDE follows the same strategy as presented in~\cite{chung2022dps}. We referred to the methods that iteratively applied projections onto convex sets (POCS) as score-SDE. Additional comparisons to  DDNM~\cite{wang2022ddnm} and DiffPIR~\cite{zhu2023DiffPIR} are also provided in SuppMat. All methods were implemented using their respective public repositories.

\begin{table}[b]
  \centering
  \setlength{\tabcolsep}{7pt}
  \caption{Quantitative results for Gaussian deblurring and random inpainting (70\%) on FFHQ dataset. Best: \textbf{bold}, second-best: \underline{underlined}. Comparison methods are omitted if they could not be implemented reliably for the given inverse problem task.}
  \vspace{-.2cm}
  \label{tab:gauss_inpaint_comparison}
  \begin{tabular}{@{}p{3cm}llllll@{}}
    \toprule
    \multirow{2}{*}{Method} & \multicolumn{3}{c}{Gaussian Deblurring} & \multicolumn{3}{c}{Random Inpainting} \\
    \arrayrulecolor{gray}\cmidrule(lr){2-4} \cmidrule(lr){5-7}
     & LPIPS$\downarrow$ & SSIM$\uparrow$ & PSNR$\uparrow$ & LPIPS$\downarrow$ & SSIM$\uparrow$ & PSNR$\uparrow$ \\
    \arrayrulecolor{black}\midrule
    DPS~\cite{chung2022dps} & \underline{0.128} & \underline{0.718} & \underline{25.20} & 0.104 & 0.811 & \textbf{28.03} \\
    MCG~\cite{chung2022mcg} & 0.558 & 0.509 & 15.12 & 0.145 & 0.754 & 25.33 \\
    $\mathrm{\Pi}$GDM~\cite{song2022pgdm} & - & - & - & \underline{0.086} & \textbf{0.842} & 26.62 \\
    DDRM~\cite{kawar2022ddrm} & 0.183 & 0.702 & 24.42 & 0.198 & 0.741 & 25.17 \\
    Score-SDE~\cite{song2020sde, choi2021ilvr, chung2022ccdf} & 0.571 & 0.496 & 15.17 & 0.224 & 0.718 & 24.44 \\
    \arrayrulecolor{gray}\midrule
    ZAPS \textbf{(Ours)} & \textbf{0.121} & \textbf{0.757} & \textbf{26.06} & \textbf{0.078} & \underline{0.813} & \underline{27.79} \\
    \arrayrulecolor{black}\bottomrule
  \end{tabular}
\end{table}

\begin{table}[tb]
  \centering
  \setlength{\tabcolsep}{7pt}
  \caption{Quantitative results for motion deblurring and super-resolution ($\times 4$) on FFHQ dataset. Best: \textbf{bold}, second-best: \underline{underlined}. Comparison methods are omitted if they could not be implemented reliably for the given inverse problem task.}
  \vspace{-.2cm}
  \label{tab:motion_sr_comparison}
  \begin{tabular}{@{}p{3cm}llllll@{}}
    \toprule
    \multirow{2}{*}{Method} & \multicolumn{3}{c}{Motion Deblurring} & \multicolumn{3}{c}{Super-Resolution ($\times 4$)} \\
    \arrayrulecolor{gray}\cmidrule(lr){2-4} \cmidrule(lr){5-7}
     & LPIPS$\downarrow$ & SSIM$\uparrow$ & PSNR$\uparrow$ & LPIPS$\downarrow$ & SSIM$\uparrow$ & PSNR$\uparrow$ \\
    \arrayrulecolor{black}\midrule
    DPS~\cite{chung2022dps} & \underline{0.143} & \underline{0.704} & \underline{24.03} & 0.168 & 0.719 & 23.86 \\
    MCG~\cite{chung2022mcg} & 0.565 & 0.497 & 15.10 & 0.229 & 0.623 & 20.74 \\
    $\mathrm{\Pi}$GDM~\cite{song2022pgdm} & - & - & - & \underline{0.131} & \underline{0.760} & 24.48 \\
    DDRM~\cite{kawar2022ddrm} & - & - & - & 0.175 & 0.711 & \underline{24.55} \\
    Score-SDE~\cite{song2020sde, choi2021ilvr, chung2022ccdf} & 0.546 & 0.488 & 15.02 & 0.257 & 0.609 & 19.13 \\
    \arrayrulecolor{gray}\midrule
    ZAPS \textbf{(Ours)} & \textbf{0.141} & \textbf{0.709} & \textbf{24.16} & \textbf{0.104} & \textbf{0.768} & \textbf{26.63} \\
    \arrayrulecolor{black}\bottomrule
  \end{tabular}
\end{table}

\begin{table}[b]
  \centering
  \setlength{\tabcolsep}{3pt}
  \caption{Computational costs of methods in terms of NFEs and wall-clock time (WCT)}
  \label{tab:nfe_time}
  \vspace{-.2cm}
  \begin{tabular}{@{}p{2cm}cccccc@{}}
    \toprule
    & DPS~\cite{chung2022dps} & MCG~\cite{chung2022mcg} & $\mathrm{\Pi}$GDM~\cite{song2022pgdm} & DDRM~\cite{kawar2022ddrm} & Score-SDE~\cite{song2020sde} & ZAPS \\
    \arrayrulecolor{black}\midrule
    Total NFEs & 1000 & 1000 & 100 & 20 & 1000 & 300 \\
    WCT (s) & 47.25 & 48.83 & 4.53 & 2.12 & 23.47 & 14.71 \\
    \arrayrulecolor{black}\bottomrule
  \end{tabular}
\end{table}

\vspace{-.2cm}
\subsubsection{Quantitative and Qualitative Results.}
We evaluated our method quantitatively using learned perceptual image patch similarity (LPIPS) distance, structural similarity index (SSIM), and peak signal-to-noise-ratio (PSNR). Representative results in \cref{fig:3} shows that DDRM yields blurry results in Gaussian deblurring task. DPS improves sharpness across these distinct inverse problem tasks, while ZAPS yields comparable sharpness while exhibiting a higher similarity to the ground truth, all within a third of the total NFEs.

Representative inpainting results in \cref{fig:4} show that ZAPS substantially improves upon DDRM, a method that uses a slightly lower 20 timesteps, and achieves better similarity to the ground truth and sharpness compared to DPS, which uses almost $33\times$ more steps. Similarly, when compared with $\mathrm{\Pi}$GDM, it is evident that our method gives comparable results even though $3-4\times$ fewer number of steps are used. The zoomed insets highlight subtle improvements afforded by our method compared to state-of-the-art DPS and $\mathrm{\Pi}$GDM, as seen around the eyes. 

\cref{tab:gauss_inpaint_comparison} and \cref{tab:motion_sr_comparison} show the three quantitative metrics for all methods, while \cref{tab:nfe_time} illustrates their computational complexity. ZAPS outperforms Score-SDE, MCG, and our baseline state-of-the-art comparison, DPS, in computational complexity and quantitative performance, yielding faster and improved reconstructions. Although DDRM and $\mathrm{\Pi}$GDM surpass ZAPS in terms of computational complexity, ZAPS outperforms both methods quantitatively in terms of all three metrics. Furthermore, $\mathrm{\Pi}$GDM could not be implemented reliably for several linear inverse problems related to deblurring. We also note that the parameters in ZAPS are adaptive, meaning one can reach the same computational complexity by adjusting total epochs or steps, in trade-off for a slight decrease in performance, as studied in \cref{sec:ablation}.

\subsection{Ablation Studies} \label{sec:ablation}

\begin{figure}[tb]
  \centering
  \begin{subfigure}{0.92\linewidth}
    \includegraphics[width=1.0\columnwidth]{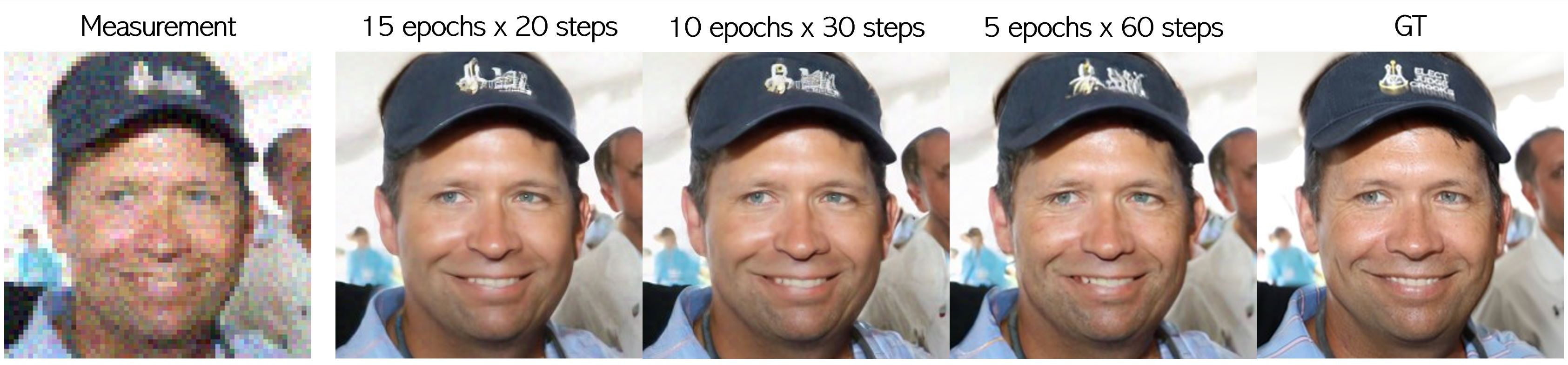}
    \caption{Reconstructions using ZAPS for super-resolution ($\times 4$) task with different total timesteps-epochs combinations for the same NFE$=300$.}
    \label{fig:5a}
  \end{subfigure}
  \hfill
  \begin{subfigure}{0.4\linewidth}
    \includegraphics[width=1.0\columnwidth]{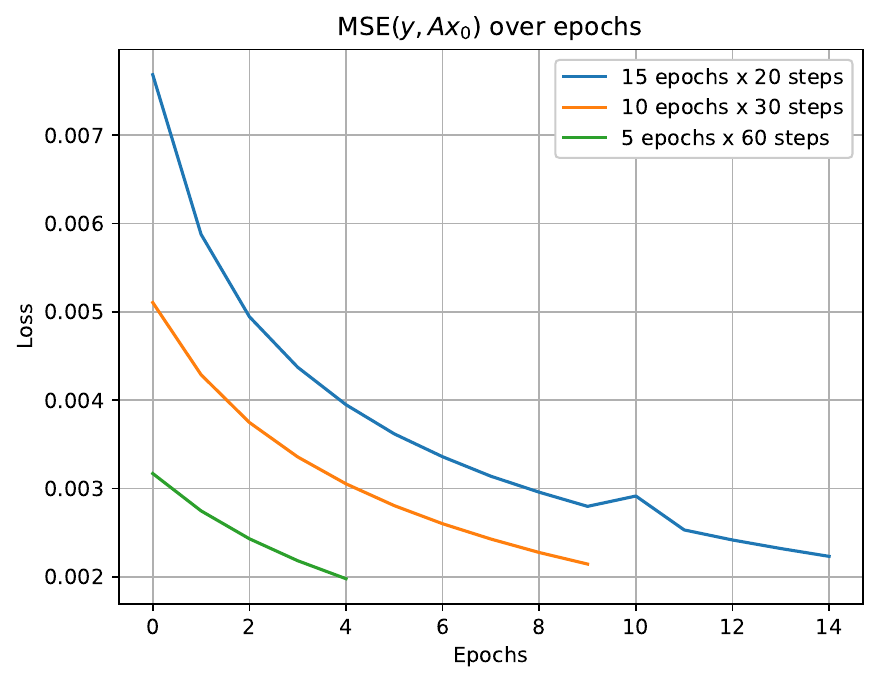}
    \caption{Loss graphs for each combination. }
    \label{fig:5b}
  \end{subfigure}
  \hspace{3.5mm}
  \begin{subfigure}{0.4\linewidth}
    \includegraphics[width=1.0\columnwidth]{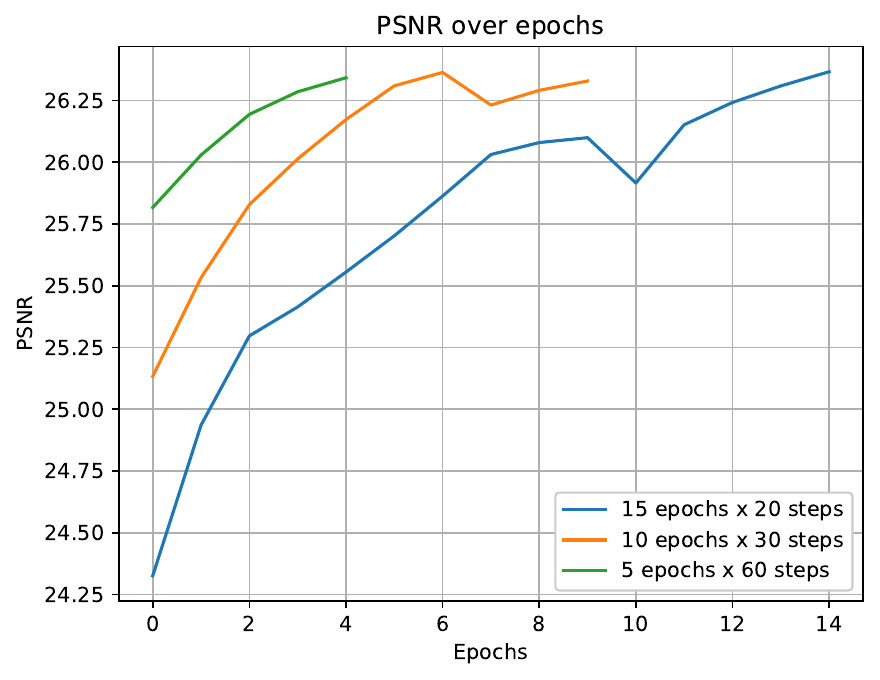}
    \caption{PSNR graphs for each combination.}
    \label{fig:5c}
  \end{subfigure}
  \caption{Study on different epochs and sampling steps combinations with fixed NFE. Results show similar quality for combinations with lower timestep approaches staring from higher loss/lower PSNR but converging to similar values.}
  \label{fig:example5}
\end{figure}

We conducted three distinct ablation studies to investigate critical aspects of our algorithm's performance. The first ablation study compared combinations of different timesteps and epochs with a fixed NFE budget, providing a nuanced exploration into the influence of specific combinations on the model's behavior. Specifically, we explored the reconstruction capabilities of the model qualitatively and quantitatively by varying the length of model timesteps, $S \in \{20,30,60\}$. For a fixed NFE budget of 300, these corresponded to 15, 10 and 5 epochs for zero-shot fine-tuning respectively. \cref{fig:5a} shows the final estimates, while \cref{fig:5b} and \cref{fig:5c} depict the corresponding loss and PSNR curves for each combination (Further quantitative results are in SuppMat). Notably, all the estimates are similar, though sharpness improves slightly as $S$ increases. However, the trade-off for choosing a high $S$ is the low number of epochs. Especially for cases, where the measurement system or noise level changes, this makes fine-tuning susceptible to initialization of the hyperparameters as it is more difficult to converge to a good solution in $\sim 5$ epochs. Thus, for improved generalizability and robustness, we opted to use $S = 30$ and 10 epochs for our database testing.

\begin{table}[b]
  \centering
  \scriptsize
  \setlength{\tabcolsep}{2.5pt}
  \caption{Quantitative results for super-resolution ($\times 4, \sigma=0.05$) on FFHQ dataset using the same NFE for each method. Best: \textbf{bold}, second-best: \underline{underlined}.}
  \label{tab:supp_3}
  \begin{tabular}{@{}p{2.5cm}lllllllll@{}}
    \toprule
    \multirow{2}{*}{Method} & \multicolumn{3}{c}{NFE=100} & \multicolumn{3}{c}{NFE=300} & \multicolumn{3}{c}{NFE=500}\\
    \arrayrulecolor{gray}\cmidrule(lr){2-4} \cmidrule(lr){5-7} \cmidrule(lr){8-10}
     & LPIPS$\downarrow$ & SSIM$\uparrow$ & PSNR$\uparrow$ & LPIPS$\downarrow$ & SSIM$\uparrow$ & PSNR$\uparrow$ & LPIPS$\downarrow$ & SSIM$\uparrow$ & PSNR$\uparrow$ \\
    \arrayrulecolor{black}\midrule
    DPS~\cite{chung2022dps} & 0.344 & 0.478 & 16.96 & 0.257 & 0.577 & 20.01 & 0.218 & 0.623 & 21.52 \\
    $\mathrm{\Pi}$GDM~\cite{song2022pgdm} & 0.131 & \textbf{0.760} & 24.48 & \underline{0.117} & \underline{0.758} & 24.80 & 0.123 & \underline{0.762} & 24.25 \\
    \arrayrulecolor{gray}\midrule
    ZAPS \textbf{(Uniform)} & \underline{0.108} & \underline{0.749} & \underline{25.92} & 0.119 & 0.729 & \underline{26.29} & \underline{0.115} & 0.756 & \underline{25.63} \\
    ZAPS \textbf{(Irregular)} & \textbf{0.106} & 0.741 & \textbf{26.08} & \textbf{0.104} & \textbf{0.768} & \textbf{26.63} & \textbf{0.095} & \textbf{0.770} & \textbf{26.26} \\
    \arrayrulecolor{black}\bottomrule
  \end{tabular}
\end{table}

Our second ablation study analyzed the performance of ZAPS with respect to other state-of-the-art methods when all methods used the same NFE. We investigated total NFEs of 100, 300, and 500 to demonstrate the robustness of our approach, given its adaptable parameters, as previously discussed. For 100 NFEs, we applied 20 steps (schedule = ``$10,7,3$'') with 5 epochs, whereas for 300 and 500 NFEs, we applied 30 steps (schedule = ``$15,10,5$'') and 50 steps (schedule = ``$30,15,5$''), respectively, for 10 epochs. Additionally, we also implemented ZAPS with uniformly spaced noise schedules to highlight the benefits of the proposed irregular noise schedules. As seen in \cref{tab:supp_3,tab:supp_3_2}, ZAPS with irregular noise schedules outperforms the state-of-the-art methods for NFE budgets of 100, 300 and 500 in super-resolution and random inpainting tasks. We note that we could not perform this test for deblurring experiments as $\mathrm{\Pi}$GDM could not be implemented reliably across the database, as previously mentioned. We also note that the difference between irregular and uniform noise schedules for ZAPS is less pronounced for 100 NFEs, but the advantage of irregular schedules becomes apparent for 300 and 500 NFEs.

\begin{table}[tb]
  \centering
  \scriptsize
  \setlength{\tabcolsep}{2.5pt}
  \caption{Quantitative results for random inpainting ($70\%, \sigma=0.05$) on FFHQ dataset using the same NFE for each method. Best: \textbf{bold}, second-best: \underline{underlined}.}
  \label{tab:supp_3_2}
  \begin{tabular}{@{}p{2.5cm}lllllllll@{}}
    \toprule
    \multirow{2}{*}{Method} & \multicolumn{3}{c}{NFE=100} & \multicolumn{3}{c}{NFE=300} & \multicolumn{3}{c}{NFE=500}\\
    \arrayrulecolor{gray}\cmidrule(lr){2-4} \cmidrule(lr){5-7} \cmidrule(lr){8-10}
     & LPIPS$\downarrow$ & SSIM$\uparrow$ & PSNR$\uparrow$ & LPIPS$\downarrow$ & SSIM$\uparrow$ & PSNR$\uparrow$ & LPIPS$\downarrow$ & SSIM$\uparrow$ & PSNR$\uparrow$ \\
    \arrayrulecolor{black}\midrule
    DPS~\cite{chung2022dps} & 0.268 & 0.593 & 20.01 & 0.189 & 0.704 & 23.74 & 0.152 & 0.754 & 25.59 \\
    $\mathrm{\Pi}$GDM~\cite{song2022pgdm} & \underline{0.086} & \textbf{0.842} & \underline{26.62} & \underline{0.080} & \textbf{0.849} & 25.06 & 0.082 & \textbf{0.845} & 24.94 \\
    \arrayrulecolor{gray}\midrule
    ZAPS \textbf{(Uniform)} & 0.122 & 0.780 & 26.20 & 0.127 & 0.773 & \underline{25.87} & \underline{0.080} & 0.791 & \underline{26.94} \\
    ZAPS \textbf{(Irregular)} & \textbf{0.085} & \underline{0.794} & \textbf{27.03} & \textbf{0.078} & \underline{0.813} & \textbf{27.79} & \textbf{0.071} & \underline{0.818} & \textbf{28.11} \\
    \arrayrulecolor{black}\bottomrule
  \end{tabular}
\end{table}

The final ablation study, exploring the benefits of using distinct weights $\zeta_t$ for each timestep versus a shared weight $\zeta$ for every step, is provided in SuppMat.

\subsection{Limitations}
The loss function we use, $\mathcal{L}(\mathbf{y},\mathbf{x}_0) = ||{\bf y} - {\bf A}{\bf x}_0||_2^2$, resembles a deep image prior-like loss~\cite{ulyanov2018dip}. However, note that there is a subtle difference in our context, where it corresponds to the log-likelihood of $p({\bf y}| {\bf x}_0)$, which is different then the (approximate) log-likelihood guidance term $p({\bf y}| {\bf x}_t)$ used at each time-step. This allows for more robustness to overfitting that is typically observed in DIP-type methods. Further overfitting avoidance measures can be taken by data-splitting~\cite{krull2019noise2void,batson2019noise2self,moran2020noisier2noise,yaman2021zero,yaman2020self}, though this was not necessary for the small number of epochs used for fine-tuning. Additionally, while our approximation in \cref{eq:hessian_approx} produces competitive results, it is important to keep in mind that wavelets may not fully decorrelate the observed Fisher information matrix. Finally, we note that while we chose DPS as a baseline for its versatility in inverse problem tasks, the adaptive weighting strategy in ZAPS, as well as our Hessian approximation, are applicable to other posterior sampling diffusion models for inverse problems.

\section{Conclusion}
In this work, we proposed a novel approach named zero-shot approximate posterior sampling (ZAPS), which harnesses zero-shot learning for dynamic automated hyperparameter tuning during the inference phase to enhance the reconstruction quality of solving linear noisy inverse problems using diffusion models. In particular, learning the log-likelihood weights facilitates the usage of more complex and irregular noise schedules, whose feasibility for inverse problems was shown, to the best of our knowledge, for the first time in this paper. These irregular noise schedules enabled high quality reconstructions with $20-50\times$ fewer timesteps. When number of epochs for fine-tuning is also considered, our approach results in a speed boost of approximately $3\times$ compared to state-of-the-art methods like DPS. Quantitative and qualitative evaluations on natural images illustrate our method's ability to attain state-of-the-art performance across diverse inverse problem tasks.

\section*{Acknowledgements}
This work was partially supported by NIH R01HL153146 and NIH R01EB032830.


%
%

\bibliographystyle{splncs04}
\bibliography{main}

\clearpage

\setcounter{section}{0}
\renewcommand\thesection{\Alph{section}}

\section{Implementation Details}
\label{sec:irregular_noise_sch}
\subsection{Irregular Noise Schedules}

\begin{figure}[tb]
  \centering
  \begin{subfigure}{\columnwidth}
    \includegraphics[width=1.0\columnwidth]{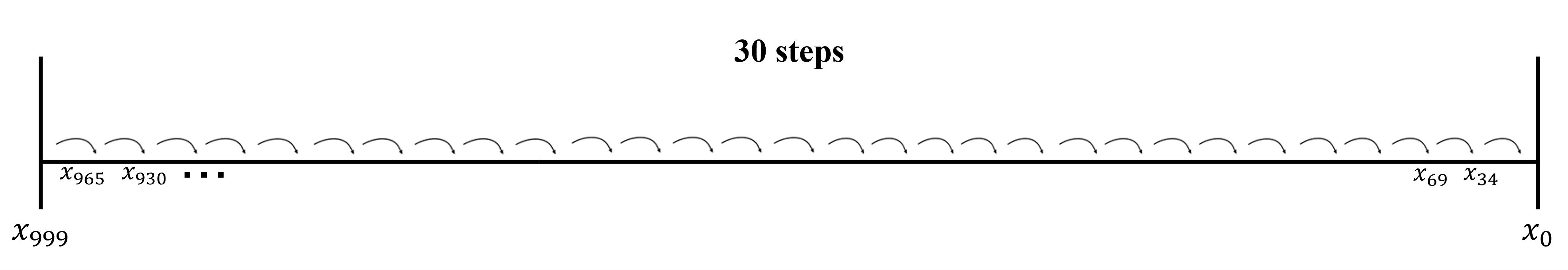}
    \caption{Fast sampling scheme that uses uniform jumps across the noise schedule.}
    \label{fig:supp_schedule_a}
  \end{subfigure}
  \hfill
  \begin{subfigure}{\columnwidth}
    \includegraphics[width=1.0\columnwidth]{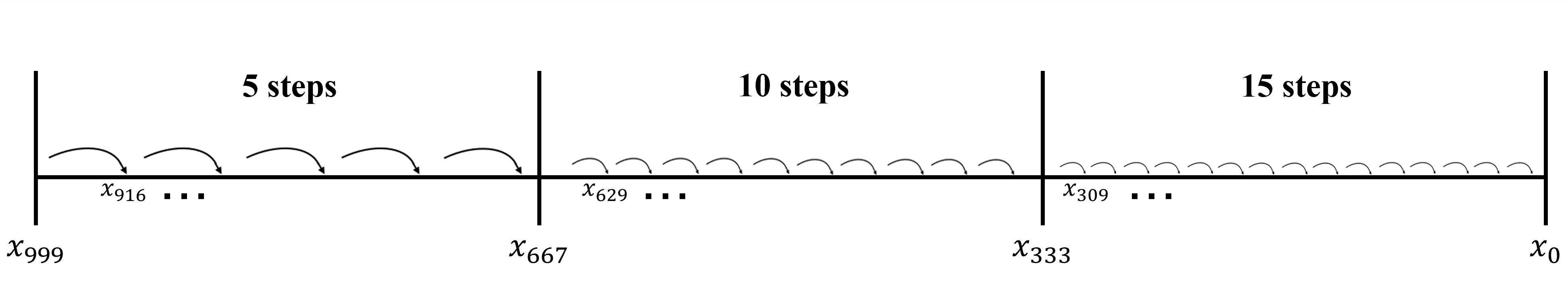}
    \caption{Fast sampling scheme that uses irregular jumps across the noise schedule.}
    \label{fig:supp_schedule_b}
  \end{subfigure}
  \caption{Illustrative figure for uniform/irregular noise schedules.}
  \label{fig:supp_schedule}
\end{figure}

 Sampling process for diffusion models can be accelerated via skipping some steps in the diffusion process~\cite{karras2022elucidating,song2020ddim,dhariwal2021diffusionsweep}. A straightforward approach is to use uniformly spaced jumps across the noise schedule (see \cref{fig:supp_schedule_a}) where the sampling path is uniformly spaced out by the selected number of steps in a regular manner. 
 A schedule we commonly use in this study is a ``$15,10,5$'' schedule, which is pictorially depicted in \cref{fig:supp_schedule_b}. This amounts to partitioning the total number of steps used in training into $3$ parts and taking uniformly spaced $5,10, \text{ and } 15$ samples from the respective segments, leading to increased sampling frequency at the lower noise levels. Although the samples are taken uniformly inside a given segment, each segment has different number of steps, making the whole schedule irregular (see \cref{fig:supp_schedule_b}). We note that this irregular noise schedule is based on the ones proposed in \cite{dhariwal2021diffusionsweep}, and the number of segments and the number of steps in each segment are chosen for the inverse problem setup based on computation time constraints while ensuring generalizability. We also note that a superior schedule may exist for a specific inverse problem, and optimization of these irregular noise schedules is an open problem to the best of our knowledge.

\subsection{Model Details}
Pre-trained models for FFHQ and ImageNet were taken from \cite{choi2021ilvr} and \cite{dhariwal2021diffusionsweep}, respectively. Both score models were used without any retraining. For our approximation for the Hessian of the log prior, we utilized Daubechies 4 (db4) wavelet as the orthogonal wavelet transform. For our database evaluation, we employed 30 timesteps with ``$15,10,5$'' schedule for 10 epochs. Furthermore, for simplicity, we opted to initialize the learnable $\{\zeta_t\}$ and $\{\mathbf{D}_t\}$ values uniformly across steps and diagonals, respectively. For $\{\zeta_t\}$ initialization in Gaussian and motion blur, $0.2$ was chosen. For random inpainting and super-resolution, $0.1$ was used. For all inverse problem tasks, diagonals of $\{\mathbf{D}_t\}$ were initialized to $0.2$. Adam optimizer with default settings was used.

\subsection{Baseline Implementations}

\subsubsection{DDRM.}
We followed the original implementation code provided by~\cite{kawar2022ddrm}, and used the default values of $\eta=0.85$ and $\eta_B=1.0$ with 20 NFE DDIM.

\subsubsection{Score-SDE, MCG, and DPS.}
For DPS implementation, we followed the original code provided by~\cite{chung2022dps}, while for MCG, we additionally performed projections onto the measurement set. For Score-SDE, we again employed projections onto the measurement set, without any gradient term to guide the diffusion process. We used 1000 NFE for each unless otherwise stated.

\subsubsection{$\mathrm{\Pi}$GDM.}
We followed the original implementation detailed in \cite{song2022pgdm} and used its public repository for implementation. We used 100 NFE unless otherwise stated.\\

All algorithms (including ZAPS) were implemented using a single NVIDIA A100 GPU. All algorithms used the same pre-trained unconditional diffusion models for a fair comparison.

\subsection{Different Sampling Strategies}
\begin{figure} [tb]
    \centering
    \includegraphics[width=1.0\columnwidth]{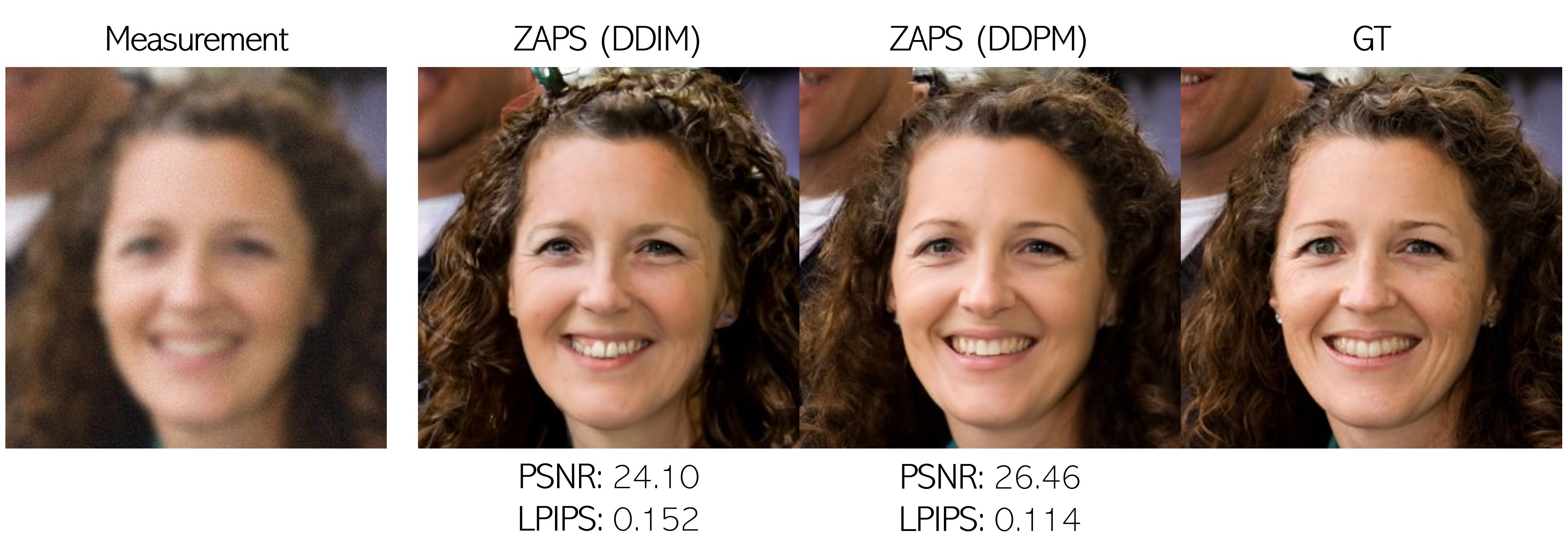}
    \caption{Representative image reconstructed with ZAPS, using DDIM and DDPM sampling schemes. DDPM exhibits superior performance to DDIM in terms of both visual quality and quantitative metrics.}
    \label{fig:supp_ddim}
\end{figure}

It is possible to use deterministic sampling schemes, such as denoising diffusion implicit models (DDIM)~\cite{song2020ddim}, to sample from a pre-trained DDPM model. Forward process for DDIM can be expressed as
\begin{equation}
    q_\sigma(\mathbf{x}_t | \mathbf{x}_{t-1}, \mathbf{x}_0) = \frac{q_\sigma(\mathbf{x}_{t-1} | \mathbf{x}_t, \mathbf{x}_0) \cdot q_\sigma(\mathbf{x}_t | \mathbf{x}_0)}{q_\sigma(\mathbf{x}_{t-1} | \mathbf{x}_0)}.
\end{equation}
As evident from observation, each $\mathbf{x}_t$ is not solely dependent on $\mathbf{x}_{t-1}$ but also on $\mathbf{x}_0$, rendering the forward process non-Markovian. Given a noisy observation $\mathbf{x}_t$, the reverse process involves initially predicting the corresponding denoised $\mathbf{x}_0$ via Tweedie's formula
\begin{equation}
    \mathbf{\hat{x}}_0 = \frac{\mathbf{x}_t - \sqrt{1 - \bar{\alpha}_t} \cdot \epsilon_\theta(\mathbf{x}_t, t)}{\sqrt{\bar{\alpha_t}}}.
\end{equation}
Using this estimate, one can generate a sample $\mathbf{x}_{t-1}$ from a sample $\mathbf{x}_t$ via:
\begin{equation}
    \mathbf{x}_{t-1} = \sqrt{\bar{\alpha}_{t-1}}\mathbf{\hat{x}}_0 + \sqrt{1 -\bar{\alpha}_{t-1} - \sigma_t^2} \cdot \epsilon_\theta(\mathbf{x}_t, t) + \sigma_t \mathbf{z},
\end{equation}
where $\sigma_t = \eta \cdot \sqrt{\frac{(1 - \bar{\alpha}_{t-1})}{(1 - \bar{\alpha}_t)} \cdot \left(1 - \frac{\bar{\alpha}_t}{\bar{\alpha}_{t-1}}\right)}$ and $\mathbf{z} \sim \mathcal{N}(\mathbf{0},\mathbf{I})$. When $\sigma_t=0$, sampling process becomes deterministic. We utilized both DDPM and DDIM with the same number of sampling steps within our ZAPS framework and as seen from \cref{fig:supp_ddim}, DDPM outperforms DDIM sampling both visually and quantitatively. Thus, we use DDPM in our work.

\section{Additional Quantitative Results}
\subsection{ImageNet Results}
\cref{tab:supp_1} depicts quantitative evaluation of the state-of-the-art methods using LPIPS, SSIM, and PSNR for noisy inverse problems ($\sigma=0.05$) on the ImageNet database. ZAPS shows competitive quantitative results either as the best or the second best among all the state-of-the-art methods.

\begin{table}[b]
  \centering
  \setlength{\tabcolsep}{7pt}
  \caption{Quantitative results for Gaussian deblurring and super-resolution ($\times 4$) on ImageNet dataset. Best: \textbf{bold}, second-best: \underline{underlined}. Comparison methods are omitted if they could not be implemented reliably for the given inverse problem task.}
  \vspace{-.2cm}
  \label{tab:supp_1}
  \begin{tabular}{@{}p{3cm}llllll@{}}
    \toprule
    \multirow{2}{*}{Method} & \multicolumn{3}{c}{Gaussian Deblurring} & \multicolumn{3}{c}{Super-Resolution ($\times 4$)} \\
    \arrayrulecolor{gray}\cmidrule(lr){2-4} \cmidrule(lr){5-7}
     & LPIPS$\downarrow$ & SSIM$\uparrow$ & PSNR$\uparrow$ & LPIPS$\downarrow$ & SSIM$\uparrow$ & PSNR$\uparrow$ \\
    \arrayrulecolor{black}\midrule
    DPS~\cite{chung2022dps} & \underline{0.230} & 0.668 & 22.16 & 0.275 & 0.673 & 21.77 \\
    MCG~\cite{chung2022mcg} & 0.317 & 0.529 & 15.25 & 0.414 & 0.397 & 15.86 \\
    $\mathrm{\Pi}$GDM~\cite{song2022pgdm} & - & - & - & \underline{0.192} & 0.707 & 22.94\\
    DDRM~\cite{kawar2022ddrm} & 0.233 & \underline{0.680} & \textbf{23.34} & 0.212 & \textbf{0.725} & \textbf{24.33} \\
    Score-SDE~\cite{song2020sde, choi2021ilvr, chung2022ccdf} & 0.324 & 0.545 & 15.41 & 0.455 & 0.361 & 14.94 \\
    \arrayrulecolor{gray}\midrule
    ZAPS \textbf{(Ours)} & \textbf{0.225} & \textbf{0.682} & \underline{22.45} & \textbf{0.186} & \underline{0.718} & \underline{23.82} \\
    \arrayrulecolor{black}\bottomrule
  \end{tabular}
\end{table}

\subsection{Comparisons with DDNM and DiffPIR}
We further compared ZAPS with the recently proposed DDNM~\cite{wang2022ddnm} and DiffPIR~\cite{zhu2023DiffPIR} for Gaussian deblurring and super-resolution ($\times 4$) tasks (see \cref{tab:tab1_rebuttal}). Each method is implemented using their respective public repository. ZAPS achieves $>20\%$ improvement in terms of LPIPS, which is a perception-oriented metric.

\begin{table}[t]
\vspace{-.5cm}
  \centering
  \footnotesize
  \setlength{\tabcolsep}{8pt}
  \caption{Quantitative results for Gaussian deblurring and super-resolution ($\times 4$) on FFHQ dataset using NFE=100 ($\sigma=0.05$) for each method. Best: \textbf{bold}, second-best: \underline{underlined}.}
  \label{tab:tab1_rebuttal}
  \begin{tabular}{@{}p{2cm}cccccc}
    \toprule
    \multirow{2.5}{*}{Method} & \multirow{2.5}{*}{NFE$\downarrow$} & \multirow{2.5}{*}{WCT(s)$\downarrow$} & \multicolumn{2}{c}{Gaussian Deblur} & \multicolumn{2}{c}{SR ($\times$ 4)} \\
    \arrayrulecolor{gray}\cmidrule(lr){4-5} \cmidrule(lr){6-7}
     & \textbf{ } & \textbf{ } & LPIPS$\downarrow$ & PSNR$\uparrow$ & LPIPS$\downarrow$ & PSNR$\uparrow$ \\
    \arrayrulecolor{black}\midrule
    DiffPIR~\cite{zhu2023DiffPIR} & 100 & 12.47 & 0.182 & 25.86 & \underline{0.143} & 26.02 \\
    DDNM~\cite{wang2022ddnm} & 100 & 11.88 & \underline{0.172} & \textbf{27.25} & 0.148 & \underline{26.76} \\
    \arrayrulecolor{gray}\midrule
    ZAPS & 100 & 10.85 & \textbf{0.128} & \underline{26.41} & \textbf{0.114} & \textbf{26.83} \\
    \arrayrulecolor{black}\bottomrule
  \end{tabular}
\end{table}

\subsection{Different Total Epochs - Timesteps Combinations with Fixed Total NFEs}
As explained in our first ablation study in the main text, we also assessed the effectiveness of various combinations of total timesteps in the posterior sampling and number of epochs for fine-tuning quantitatively (see \cref{tab:supp_4}) while keeping the NFE constant. As anticipated, decreasing the number of epochs to 5 to allocate more timesteps had an adverse impact, where for some measurements, log-likelihood weights and approximated diagonals did not have the time to stabilize during the fine-tuning. Also as expected, 15 epochs $\times$ 20 timesteps combination and 10 epochs $\times$ 30 timesteps had similar outcomes, in which the latter outperformed the former slightly, and was used in the study.

\begin{table}[b]
  \centering
  \setlength{\tabcolsep}{9pt}
  \caption{Quantitative results for different epochs-steps combination (fixed NFE$=300$) for the super-resolution ($\times 4, \sigma=0.05$) inverse problem task on FFHQ dataset. Best: \textbf{bold}, second-best: \underline{underlined}.}
  \vspace{-.2cm}
  \label{tab:supp_4}
  \begin{tabular}{@{}p{4cm}cccc@{}}
    \toprule
    Combination & Schedule & LPIPS$\downarrow$ & SSIM$\uparrow$ & PSNR$\uparrow$ \\
    \arrayrulecolor{black}\midrule
    $15$ epochs $\times$ $20$ timesteps & ``$10,7,3$'' & \textbf{0.096} & \underline{0.763} & \underline{26.48} \\
    $10$ epochs $\times$ $30$ timesteps & ``$15,10,5$'' & \underline{0.104} & \textbf{0.768} & \textbf{26.63} \\
    $\>\: 5$ epochs $\times$ $60$ timesteps & ``$30,20,10$'' & 0.109 & 0.741 & 26.39 \\
    \arrayrulecolor{black}\bottomrule
  \end{tabular}
\end{table}

\section{Additional Qualitative Results}

\subsection{Effect of Using Distinct Weights $\{\zeta_t\}$}

\begin{figure}[t]
  \centering
  \begin{subfigure}{0.8\linewidth}
    \includegraphics[width=1.0\columnwidth]{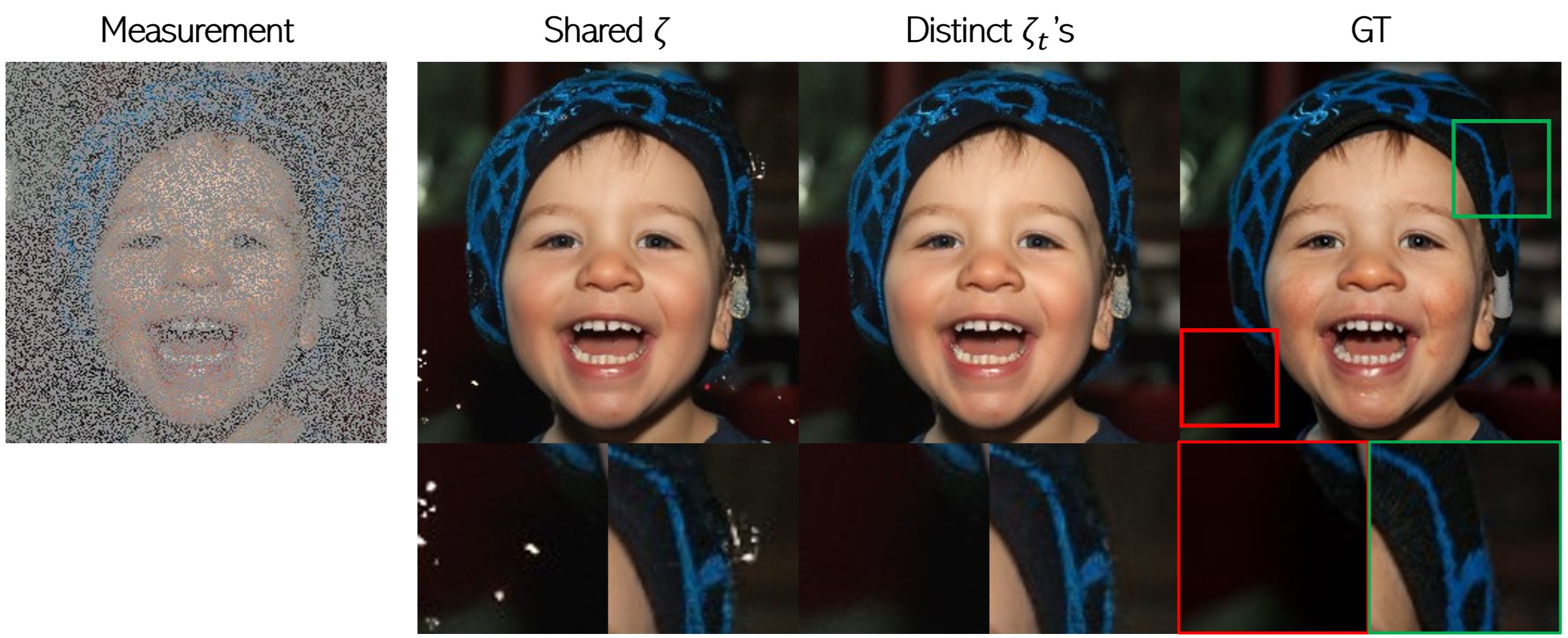}
    \caption{Results for shared and distinct log-likelihood weights for irregular timesteps.}
    \label{fig:6a}
  \end{subfigure}
  \hfill
  \begin{subfigure}{0.4\linewidth}
    \includegraphics[width=1.0\columnwidth]{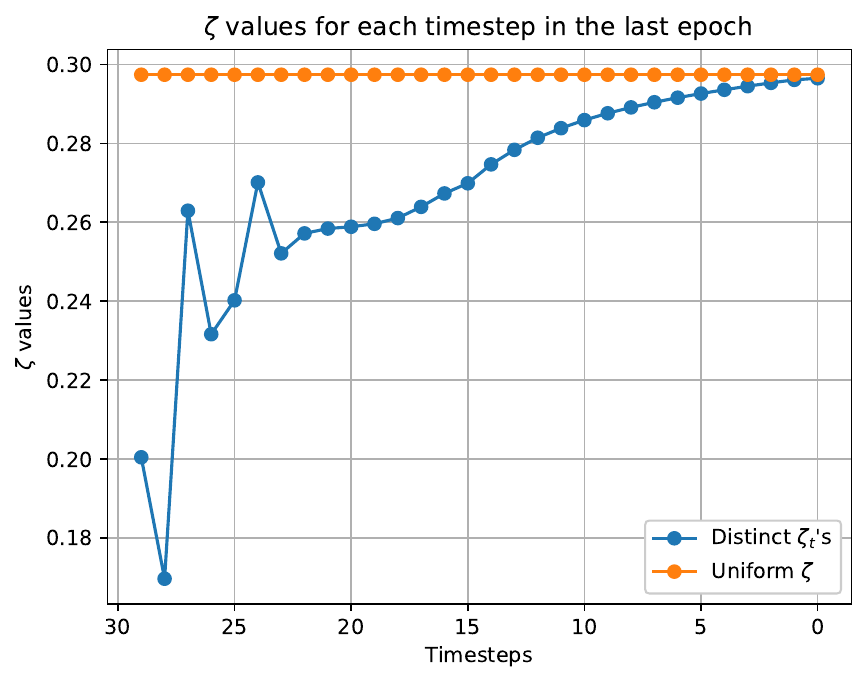}
    \caption{Distinct $\zeta_t$'s recover from sub-optimal initialization.}
    \label{fig:6b}
  \end{subfigure}
  \hspace{3.5mm}
  \begin{subfigure}{0.4\linewidth}
    \includegraphics[width=1.0\columnwidth]{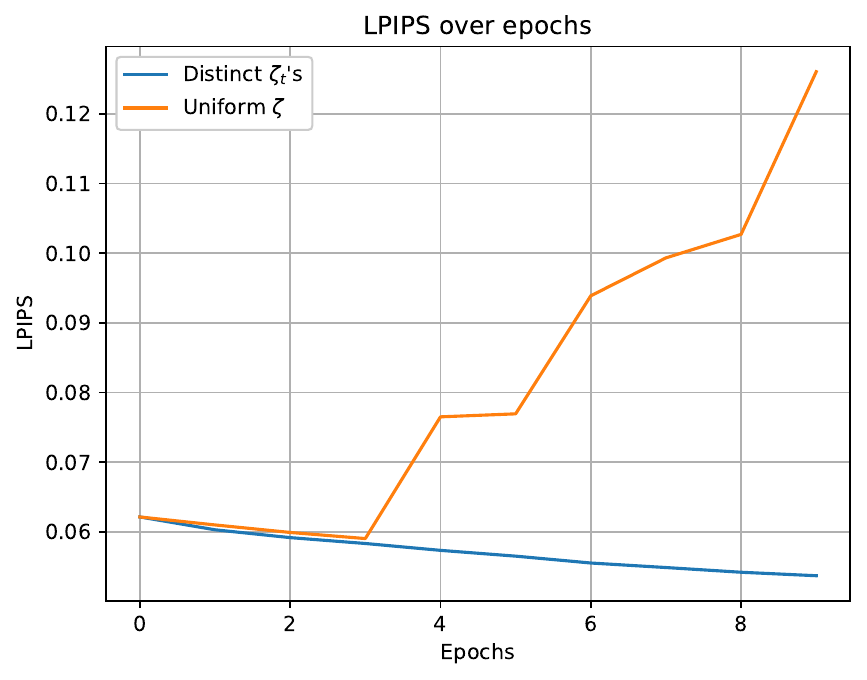}
    \caption{Shared $\zeta$ is prone to over-fitting.\\ \textbf{ }}
    \label{fig:6c}
  \end{subfigure}
  \caption{Study on different $\zeta$ choice strategies.}
  \label{fig:example6}
 \end{figure}

As part of our ablation studies, we examined the influence of selecting a shared weight $\zeta$ for every step versus using distinct weights $\zeta_t$ for each timestep. \cref{fig:6a} shows that the shared $\zeta$ approach leads to artifacts that are highlighted in the zoomed-in insets. Furthermore, \cref{fig:6b} shows that the shared approach is susceptible to the goodness of the initialization, while the adaptive $\zeta_t$ weights are able to recover from arbitrary initializations. \cref{fig:6c} further shows that the shared approach is prone to overfitting. Thus, the proposed approach with adaptive $\zeta_t$ log-likelihood weights is preferred.

\begin{figure} [tb]
    \centering
    \includegraphics[width=\columnwidth]{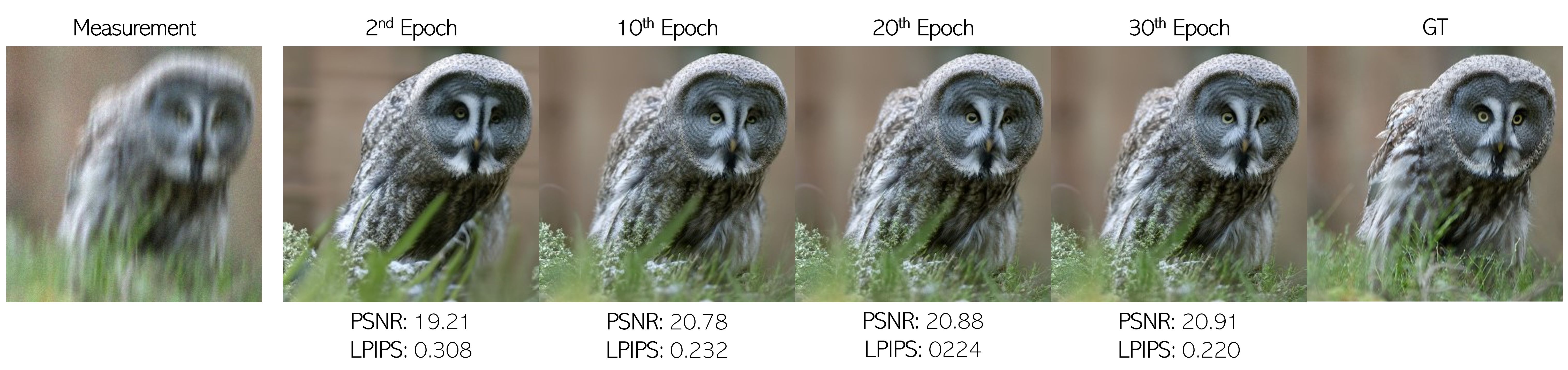}
    \caption{Different epochs ZAPS reconstruction for motion deblurring task ($\sigma=0.05$). Fine-tuning becomes redundant after the $10^{\textrm{th}}$ epoch when 30 sampling steps is being used.}
    \label{fig:supp_further_epochs}
\end{figure}

\subsection{Effect of Higher Number of Epochs for Fine-Tuning in ZAPS}
We evaluated our method on a representative ImageNet sample over $30$ epochs for motion deblurring inverse problem task using 30 steps. As seen from \cref{fig:supp_further_epochs}, both the reconstruction faithfulness, and the visual quality, measured by PSNR and LPIPS respectively, demonstrate an increase via fine-tuning. However, after the $10^{\mathrm{th}}$ epoch, the performance saturates and the gain through the log-likelihood weight update is diminished. Furthermore, no DIP-like overfitting was observed owing to the differences between the training loss function and the log-likelihood update term, as discussed in the main text. Thus, $10$ epochs were used as the maximum number of epochs for the 30 step ZAPS setting to minimize total NFEs.

\subsection{Effect of Wavelet Transform Choice}
\begin{figure} [b]
    \centering
    \includegraphics[width=\columnwidth]{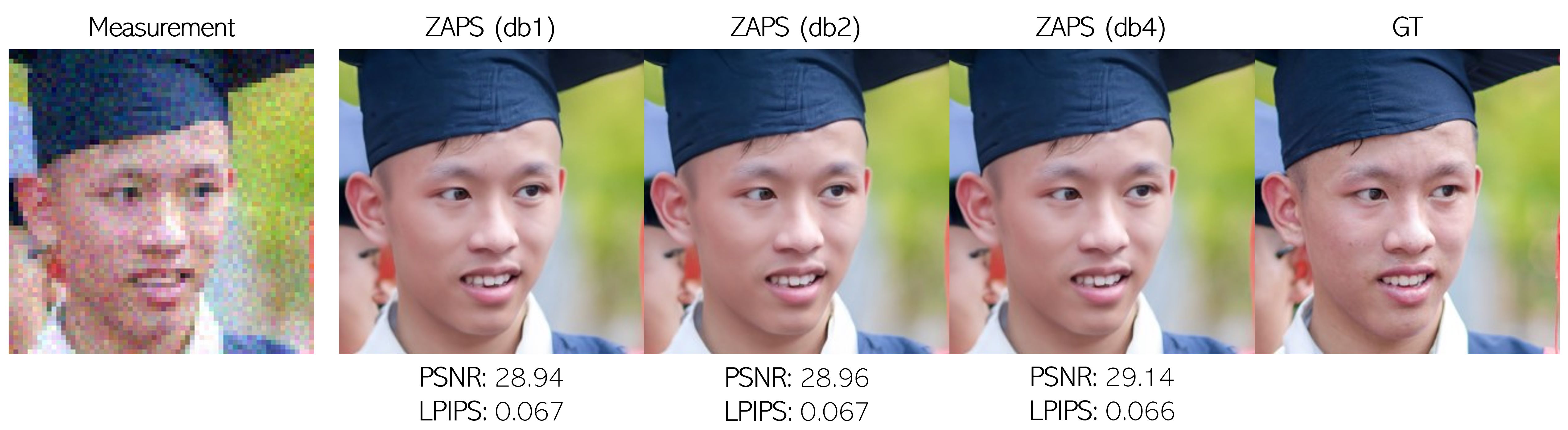}
    \caption{Illustrative ZAPS results when different orthogonal wavelets are considered. The effect of the wavelet choice is trivial as each of them can converge to a good reconstruction. As it is commonly used in practice, we decide on using Daubechies 4 wavelet.}
    \label{fig:supp_different_wavelets}
\end{figure}

We further investigated using different types of orthogonal wavelets from the Daubechies wavelet family in \cref{fig:supp_different_wavelets}. As seen from the results, the effect of the wavelet selection is negligible. Therefore, we opted to use Daubechies 4 wavelet as it is commonly used in sparse signal processing literature~\cite{lustig2007sparse}.

\subsection{Additional Experimental Results}
\begin{figure} [tb]
    \centering
    \includegraphics[width=0.8\columnwidth]{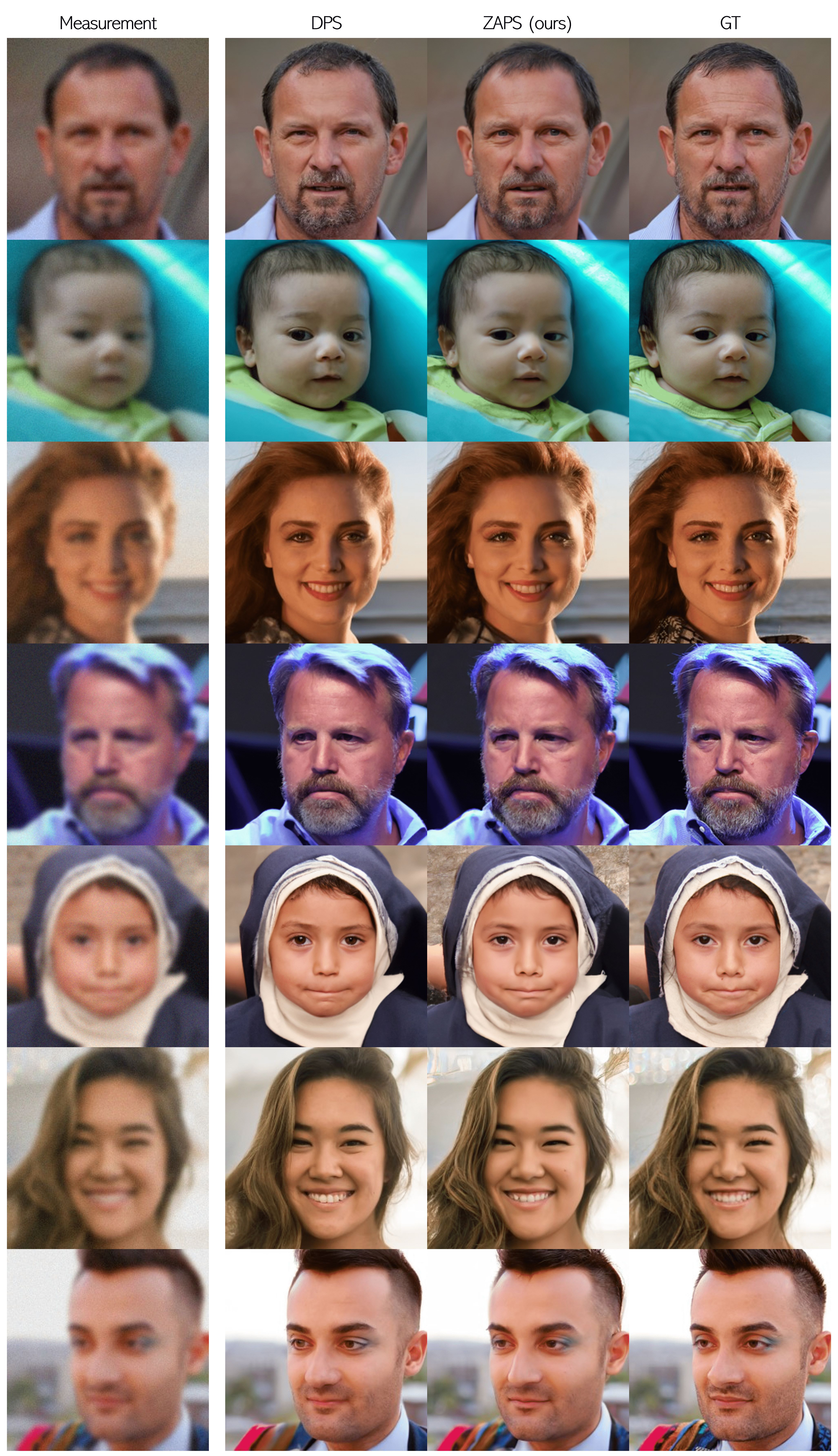}
    \caption{Gaussian deblurring results for ZAPS and DPS on FFHQ \cite{Karras2019ffhq} 256$\times$256 dataset.}
    \label{fig:supp_1}
\end{figure}

\begin{figure} [tb]
    \centering
    \includegraphics[width=0.8\columnwidth]{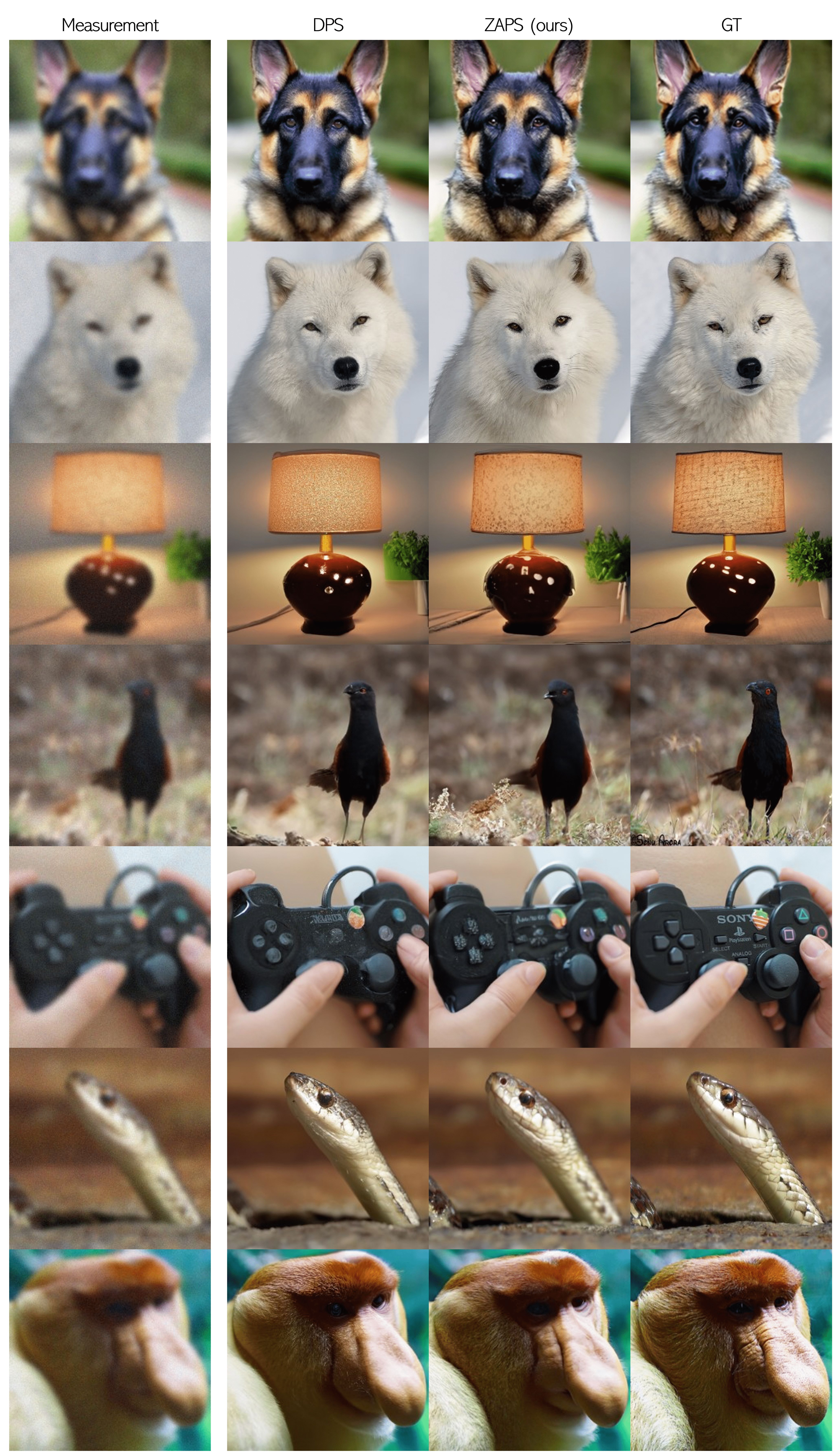}
    \caption{Gaussian deblurring results for ZAPS and DPS on ImageNet \cite{deng2009imagenet} 256$\times$256 dataset.}
    \label{fig:supp_2}
\end{figure}

\begin{figure} [tb]
    \centering
    \includegraphics[width=0.8\columnwidth]{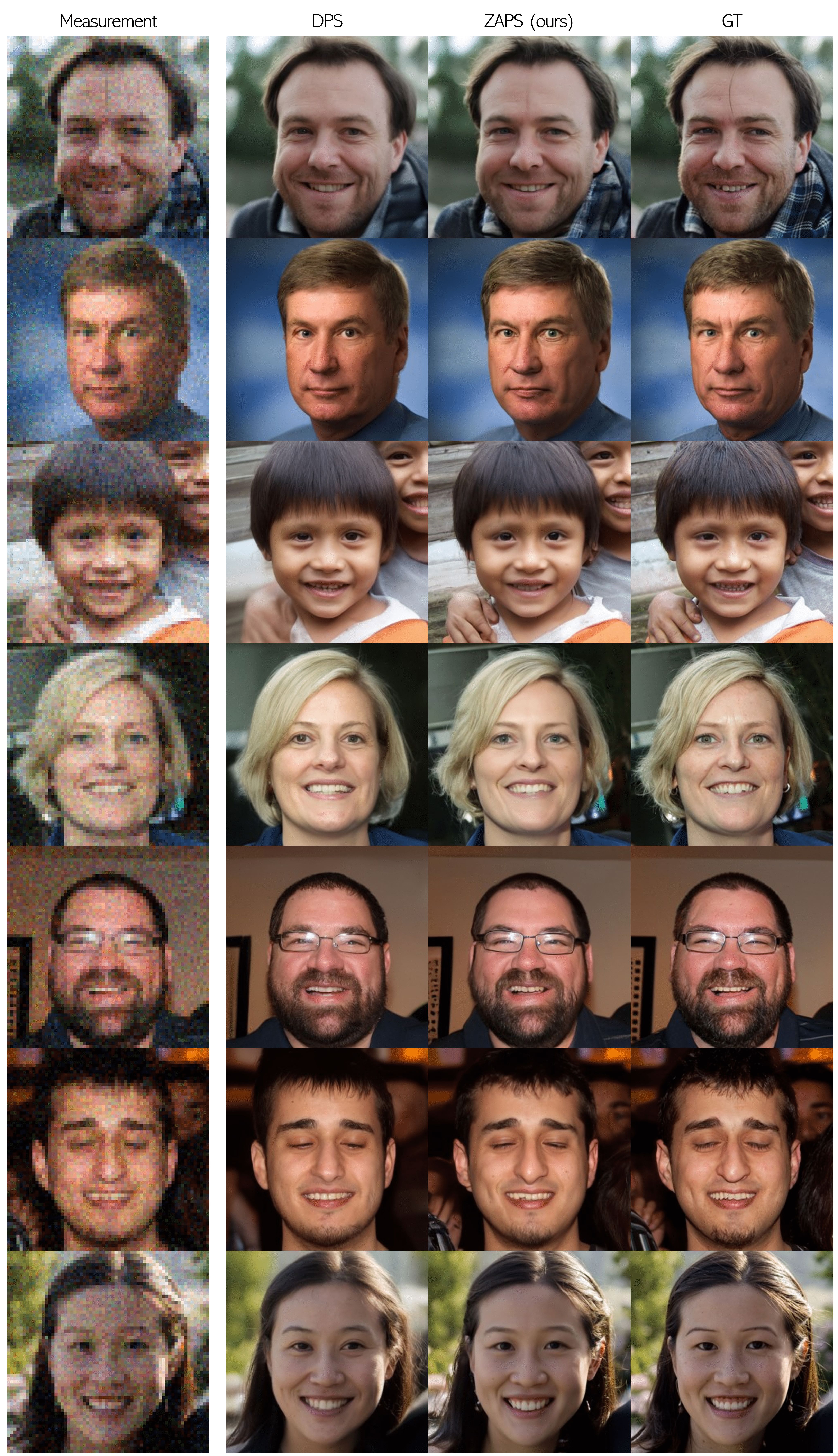}
    \caption{Super-resolution ($\times$4) results for ZAPS and DPS on FFHQ \cite{Karras2019ffhq} 256$\times$256 dataset.}
    \label{fig:supp_3}
\end{figure}

\begin{figure} [tb]
    \centering
    \includegraphics[width=0.8\columnwidth]{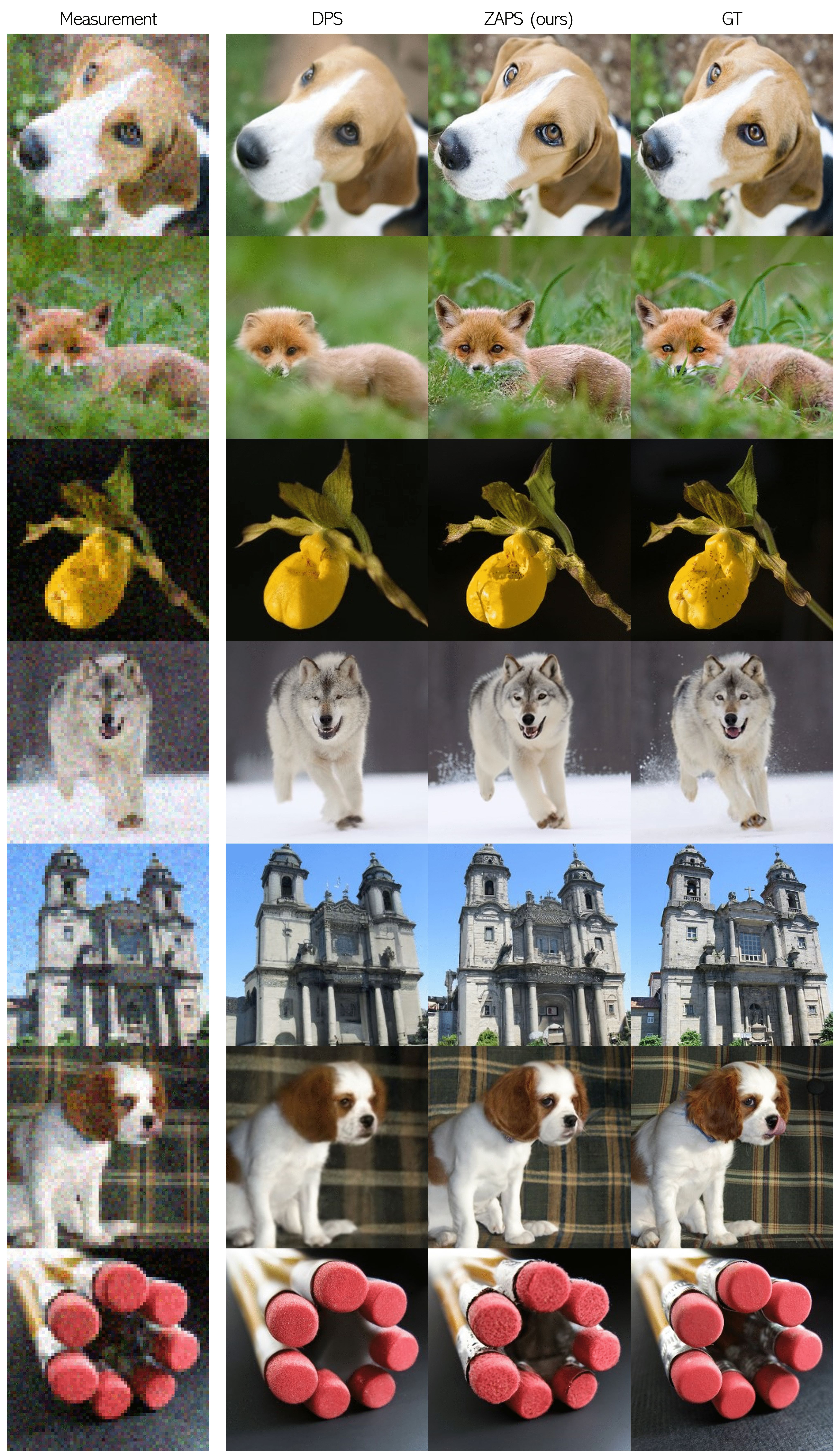}
    \caption{Super-resolution ($\times$4) results for ZAPS and DPS on ImageNet \cite{deng2009imagenet} 256$\times$256 dataset.}
    \label{fig:supp_4}
\end{figure}

\begin{figure} [tb]
    \centering
    \includegraphics[width=0.8\columnwidth]{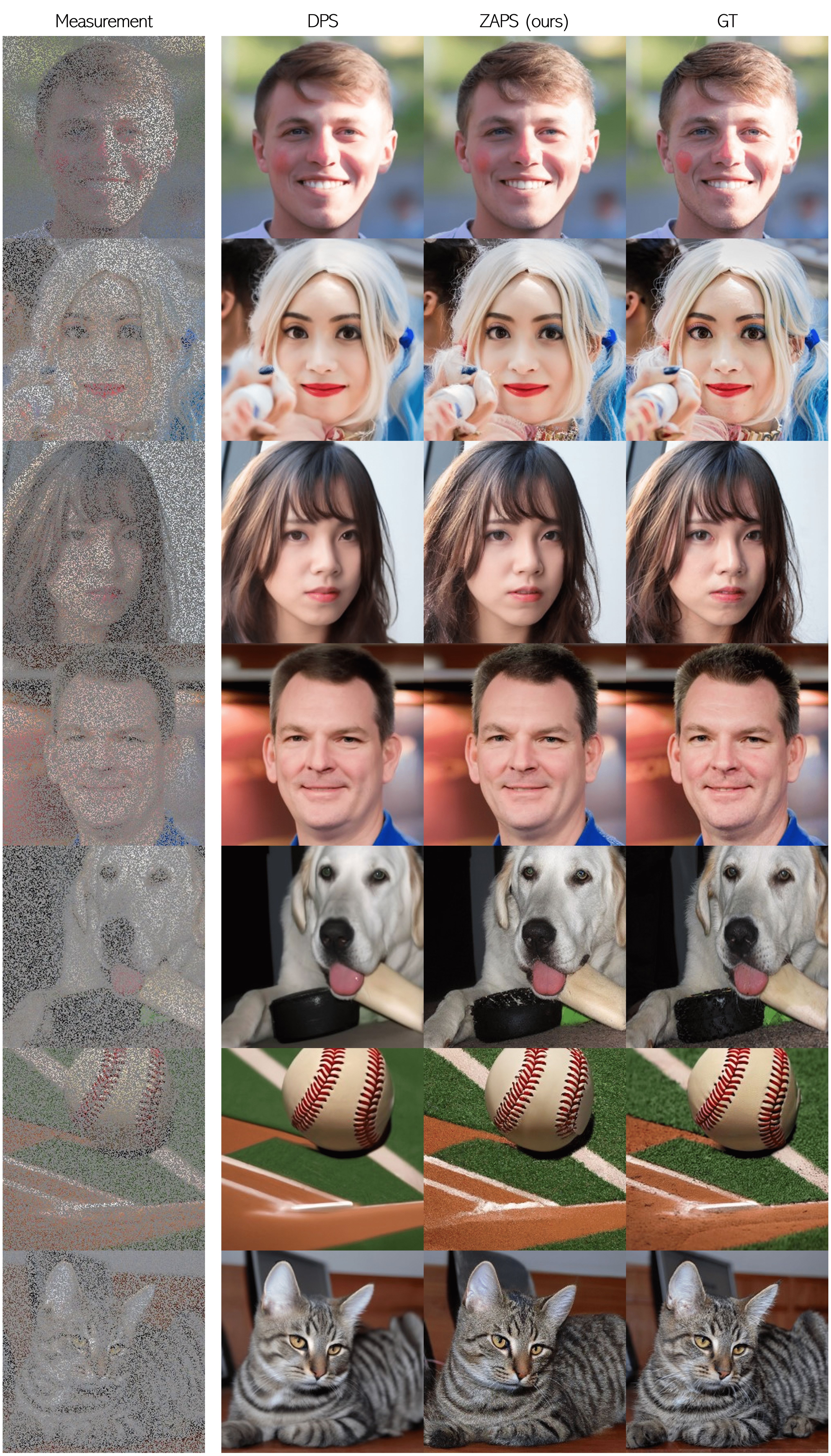}
    \caption{Random inpainting ($70\%$) results for ZAPS and DPS on FFHQ \cite{Karras2019ffhq} 256$\times$256 and ImageNet \cite{deng2009imagenet} 256$\times$256 dataset.}
    \label{fig:supp_5}
\end{figure}

\begin{figure} [tb]
    \centering
    \includegraphics[width=0.8\columnwidth]{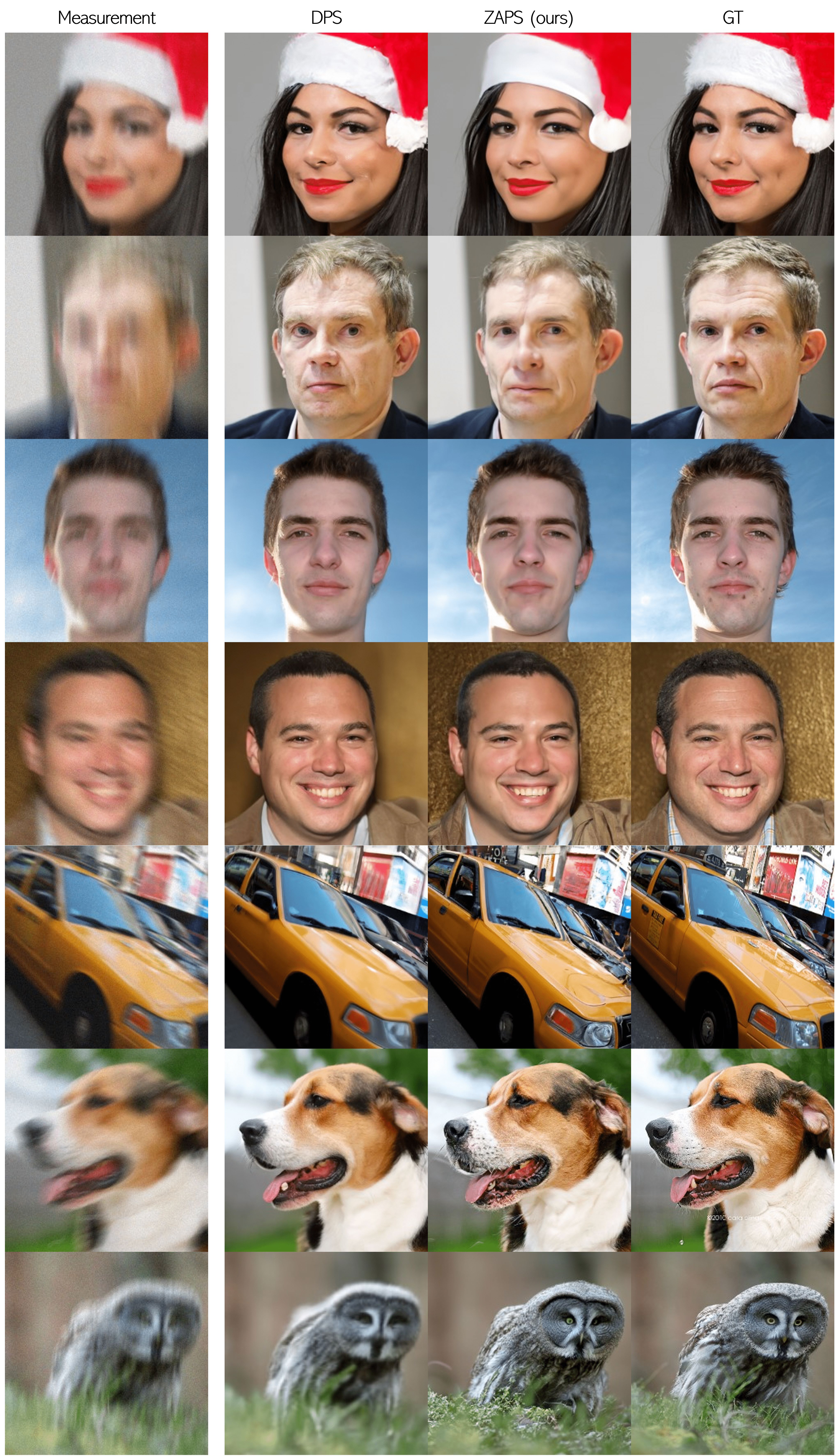}
    \caption{Motion deblurring results for ZAPS and DPS on FFHQ \cite{Karras2019ffhq} 256$\times$256 and ImageNet \cite{deng2009imagenet} 256$\times$256 dataset.}
    \label{fig:supp_6}
\end{figure}

\begin{figure} [tb]
    \centering
    \includegraphics[width=\columnwidth]{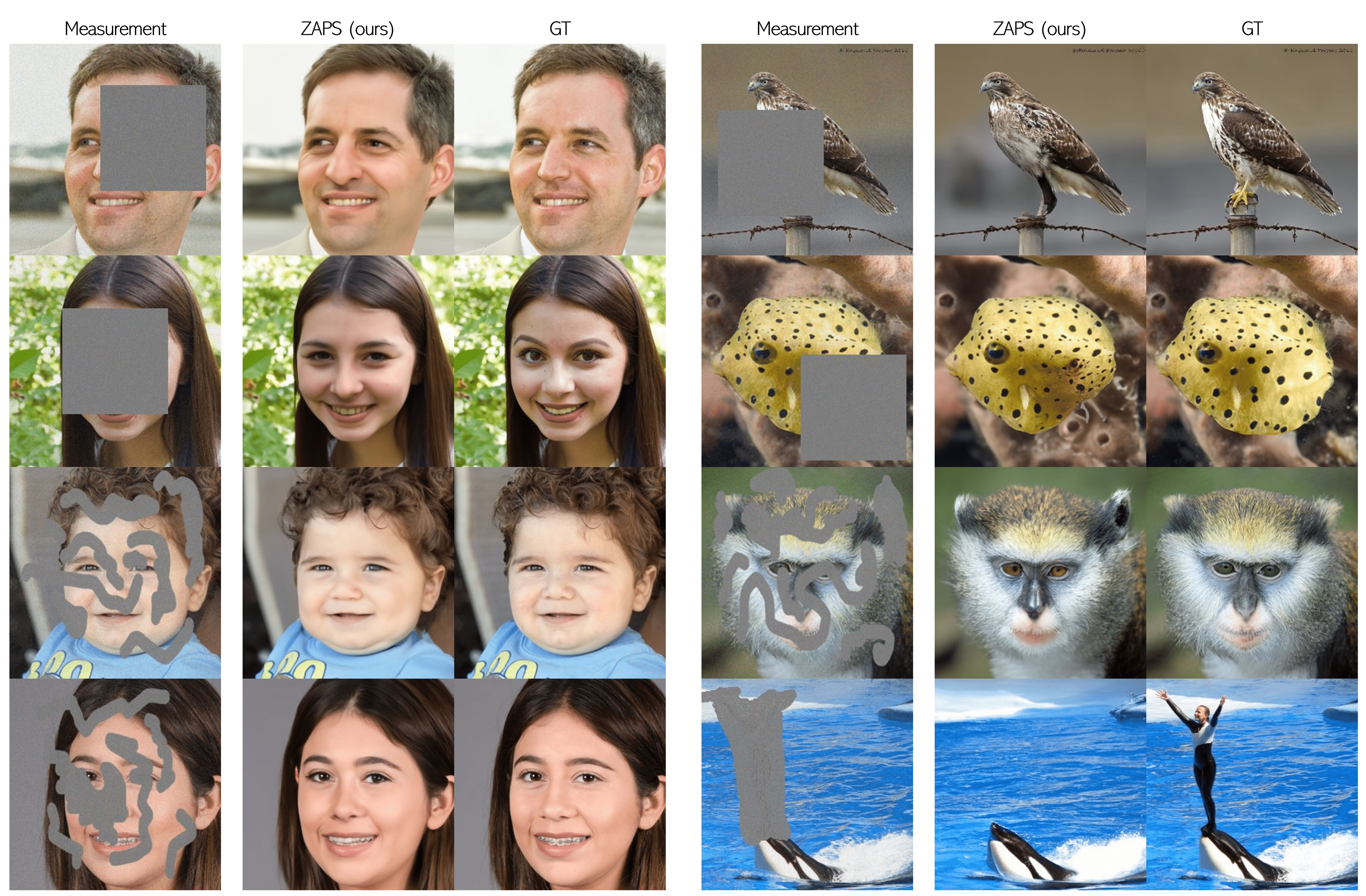}
    \caption{Representative ZAPS reconstructions for image inpainting task using different masks (box size is $128 \times 128$) with $\sigma=0.05$.}
    \label{fig:supp_different_masks}
\end{figure}

 Further qualitative experimental results, comparing ZAPS with our state-of-the-art baseline, DPS, for various noisy inverse problem tasks ($\sigma =0.05$) are given in \cref{fig:supp_1,fig:supp_2,fig:supp_3,fig:supp_4,fig:supp_5,fig:supp_6}. We also provide inpainting task outcomes in \cref{fig:supp_different_masks} for various types of masks in addition to random and rectangular box inpainting.  
 Our approach, involving the adjustment of the log-likelihood weights during fine-tuning and integration of irregular noise schedules with fewer sampling steps, results in a notable acceleration of approximately $\times 3$ on the FFHQ dataset and around $\times 4$ on ImageNet dataset, while also delivering superior performance.

\section{Derivation of $\mathrm{\Pi}$GDM Update Using Woodbury Matrix Identity}

For the log-likelihood estimation, $\mathrm{\Pi}$GDM~\cite{song2022pgdm} uses a Gaussian centered around $\mathbf{A\hat{x}}_0$ to obtain the following score approximation:
\begin{equation}
	\nabla_{\mathbf{x}_t} \log p_t(\mathbf{y}|\mathbf{x}_t) \simeq  \frac{\partial \hat{\mathbf{x}}_0}{\partial \mathbf{x}_t} \mathbf{A}^\top (r_t^2 \mathbf{A} \mathbf{A}^\top + \sigma_y^2 \mathbf{I})^{-1} (\mathbf{y} - \mathbf{A}\hat{\mathbf{x}}_0). \label{eq:supp1}
\end{equation}
In the text, we state that using Woodbury matrix identity, this can be rewritten as
\begin{equation}
    \nabla_{\mathbf{x}_t} \log p_t(\mathbf{y}|\mathbf{x}_t) \simeq \frac{\partial \hat{\mathbf{x}}_0}{\partial \mathbf{x}_t} (\mathbf{A}^\top \mathbf{A} + \eta \mathbf{I})^{-1} \mathbf{A}^\top (\mathbf{y}-\mathbf{A}\hat{\mathbf{x}}_0), \quad \text{where } \eta=\frac{\sigma_y^2}{r_t^2}, \label{eq:supp2}
\end{equation}
which is more similar in form to DPS.\newline

\noindent \textit{Proof.} One can write \cref{eq:supp1} as
\begin{equation}
    \frac{\partial \hat{\mathbf{x}}_0}{\partial \mathbf{x}_t} \mathbf{A}^\top (r_t^2 \mathbf{A} \mathbf{A}^\top + \sigma_y^2 \mathbf{I})^{-1} (\mathbf{y} - \mathbf{A}\hat{\mathbf{x}}_0) = \frac{\partial \hat{\mathbf{x}}_0}{\partial \mathbf{x}_t} \mathbf{A}^\top \frac{1}{r_t^2} (\mathbf{AA}^\top + \eta \mathbf{I})^{-1} (\mathbf{y} - \mathbf{A}\hat{\mathbf{x}}_0), \label{eq:supp3}
\end{equation}
where $\eta = \frac{\sigma_y^2}{r_t^2}$. Applying Woodbury matrix identity, $(\mathbf{AA}^\top + \eta \mathbf{I})^{-1}$ can be rewritten as

\begin{align}
    (\mathbf{AA}^\top + \eta \mathbf{I})^{-1} &= \frac{\mathbf{I}}{\eta} - \frac{\mathbf{A}}{\eta}\left(\mathbf{I} + \frac{1}{\eta} \mathbf{A}^\top \mathbf{A} \right)^{-1} \frac{\mathbf{A}^\top}{\eta}\\
    &= \frac{1}{\eta} \left( \mathbf{I} - \mathbf{A}\left(\eta\mathbf{I} + \mathbf{A}^\top \mathbf{A} \right) ^{-1} \mathbf{A}^\top \right).
\end{align}
Thus, \cref{eq:supp3} becomes
\begin{equation}
     \nabla_{\mathbf{x}_t} \log p_t(\mathbf{y}|\mathbf{x}_t) \propto \frac{\partial \hat{\mathbf{x}}_0}{\partial \mathbf{x}_t} \frac{\mathbf{A}^\top}{\eta} \left( \mathbf{I} - \mathbf{A}\left(\eta\mathbf{I} + \mathbf{A}^\top \mathbf{A} \right) ^{-1} \mathbf{A}^\top \right) (\mathbf{y} - \mathbf{A}\hat{\mathbf{x}}_0).
\end{equation}
Noting $\mathbf{I} = (\mathbf{A}^\top \mathbf{A}+\eta\mathbf{I})(\mathbf{A}^\top \mathbf{A}+\eta\mathbf{I})^{-1}$ yields
\begin{align}
    \nabla_{\mathbf{x}_t} \log p_t(\mathbf{y}|\mathbf{x}_t) &\propto \frac{\partial \hat{\mathbf{x}}_0}{\partial \mathbf{x}_t}  \mathbf{I} (\mathbf{A}^\top \mathbf{A}+\eta\mathbf{I})^{-1} \mathbf{A}^\top(\mathbf{y} - \mathbf{A}\hat{\mathbf{x}}_0),
\end{align}
which is similar to the DPS update, where $(\mathbf{A}^\top \mathbf{A}+\eta\mathbf{I})^{-1}$ is the difference.

\end{document}